\documentclass[11pt]{article}
\usepackage[T1]{fontenc}
\usepackage[utf8]{inputenc}
\usepackage{lmodern}
\usepackage[a4paper,margin=2.3cm]{geometry}
\usepackage{graphicx}
\usepackage{booktabs,tabularx,array}
\usepackage{amsmath}
\usepackage{microtype}
\usepackage{needspace}
\usepackage[hidelinks]{hyperref}
\usepackage[numbers,super,sort&compress]{natbib}
\title{Autonomous Optimization of Numerical Algorithms\\ by an AI Agent: Application to Atomistic Simulation}
\author{Jeheon Woo\\ \small Center for Artificial Intelligence and Natural Sciences,\\ Korea Institute for Advanced Study,\\ 85 Hoegiro, Dongdaemun-gu, Seoul 02455, Republic of Korea\\ \small \texttt{jhwoo1905@kias.re.kr}}
\date{}

\begin{document}
\maketitle

\begin{abstract}
Coding agents built on large language models, run in automated improvement loops, can discover more efficient algorithms. We present AIOA, a framework that brings this approach to the numerical algorithms of atomistic simulation: an agent improves an algorithm against a pre-specified metric in a loop guarded by verifiers it cannot alter. Applied to ASE, a widely used atomistic simulation library, AIOA targets two facets of an algorithm: its implementation, rewritten to run faster with unchanged results, and the method itself, redesigned to converge with fewer force evaluations. Agents made six kernels 22--24 times faster on average in Python with NumPy and SciPy. On unseen problems, optimizers from the best configuration needed 3.1--3.4 times fewer force evaluations than L-BFGS for structure relaxation and 4.6--17 times fewer than standard climbing-image nudged elastic band calculations for reaction paths. The agents' optimizers also contain two algorithmic components we did not find in the literature.
\end{abstract}

Atomistic simulation rests on a small set of numerical algorithms shared by a vast number of calculations\cite{AllenTildesley2017}. Their cost has two sources: the implementation, which sets how fast a given computation runs, and the method itself, which sets how much computation is needed, for example how many force evaluations a structure relaxation\cite{LiuNocedal1989,Bitzek2006}, a reaction-path search\cite{Henkelman2000} or a molecular-dynamics run\cite{Verlet1967} requires. Domain experts have refined both by hand for decades\cite{ASE,Packwood2016,Makri2019}, and any improvement propagates to every calculation downstream.

Coding agents built on large language models could now take on part of this refinement. Placed in automated loops that propose code changes, measure them and keep the improvements, such agents have found more efficient matrix-multiplication schemes\cite{AlphaEvolve} and sped up numerical kernels 1.3- to 2.8-fold when coupled to a compiler or profiler\cite{ComPilot,Astra,PRAGMA}. Karpathy's autoresearch\cite{autoresearch} reduces such a loop to its essentials: an agent edits a training script and keeps the edit only if the validation loss falls. On training code Karpathy had already tuned by hand, the edits cut training time by almost a fifth\cite{nanochat}. Two recent studies bring this kind of search to atomistic simulation: an evolutionary agent that refined transition-state-search workflows built on the Atomic Simulation Environment (ASE) to raise their success rate on a target reaction dataset\cite{Meissner2026}, and an autoresearch loop that rewrote the Sella optimizer\cite{Sella} for molecular relaxation under validity and generalization gates\cite{AutoSella}.

The difficulty is that an agent optimizes the number it is scored on, not the scientific result that number stands for. A faster kernel that returns subtly wrong energies silently corrupts every calculation downstream, and an optimizer tuned to the benchmark's inputs can fail on the next problem. Controlled studies of coding agents find them hardcoding expected answers, overwriting timers or tampering with scoring code in a few percent of narrow tasks\cite{RHB,RHCO} and in about a third of open-ended research runs\cite{METR2025,RHCO}. Even the system behind the matrix-multiplication results exploited the numerical integration in the scoring function it was given\cite{Georgiev2025}. In autoresearch and other loops, pass rates on held-out tests can trail those on visible tests by tens of percentage points\cite{SpecBench}. Proposed mitigations, such as hardened environments\cite{RHB}, hidden tests with trajectory audits\cite{PerfOptBench} and metrics kept outside the agent's control\cite{RHCO}, each close one route. Instructions alone barely help\cite{METR2025,BaitBench}. For autonomous optimization to serve science, its gains must hold on unseen problems, recur across independent runs and survive an audit of the checks that admitted them.

Here we present AIOA (Autonomous Iterative Optimization of Algorithms), a framework for the autonomous optimization of numerical algorithms, and apply it to those in ASE\cite{ASE}, a widely used open-source library for atomistic simulation. We target two complementary facets of a numerical algorithm: its implementation, which the agent speeds up while the results stay fixed, and the method itself, which the agent may change so that it needs fewer force evaluations. We tested the first on six numerical kernels and the second on optimizers for structure relaxation and reaction paths, starting from ASE's own methods and from scratch. The search is autoresearch's keep-or-discard loop with verification added to the keep test. What defines AIOA is the machinery around it (Fig. 1): safeguards that each close one way to game a result, a blind held-out set evaluated only after each optimizer run, and post hoc checks of what passed them, including audits of the rules that only the agent's instructions impose.

In three independent runs, agents made the six kernels 22--24-fold faster (geometric mean over kernels) in Python with NumPy and SciPy. On blind held-out problems, optimizers from two search protocols, three runs each, needed on average 1.8-fold fewer force evaluations than ASE's reference method for each task. Optimizers from the best configuration, Claude Fable 5.1 runs that could read the record of earlier runs, needed 3.1--3.4-fold fewer than L-BFGS for structure relaxation and 4.6--17-fold fewer than a standard climbing-image nudged elastic band calculation for reaction paths (Figs. 2 and 3). We also report what the safeguards caught and missed, and what the agents built, separating tuning from algorithmic change (Fig. 4) and including two components we did not find in the literature. We then ask what a number from autonomous optimization means when independent runs reach the same aggregate by different code, and how the results change with the model (Fig. 5).

\section*{Results}
\needspace{10\baselineskip}\subsection*{AIOA and the experimental design}
AIOA runs an agent in a keep-or-discard loop over a git repository (Fig. 1a). Each iteration commits one change and measures it on an optimization set. The change is kept only if the metric improves and every gate passes; a reverted change stays in the history the agent reads. Held-out problems (Table 1) and their evaluation code live outside the agent's repository (Methods).

We applied AIOA to ASE on two complementary facets of a numerical algorithm (Table 1). In optimizing the implementation, the agent sped up six numerical kernels without changing their results. The kernels were chosen as distinct kinds of computation whose stock code leaves room for speedup (Methods). The metric is wall time. The gate requires every output to match stock ASE to within floating-point reordering and the module's subset of ASE's own unit tests to pass, and an adversarial reviewer agent tries to find, by execution, an input on which the change diverges. Claude Opus 5 made three independent runs. In each, a coordinating agent ran the six kernel agents in parallel, up to 200 iterations each, and merged their branches into one tree.

In optimizing the method itself, the agent changes what is computed. An optimizer may take different steps and reach a different valid answer, so correctness becomes convergence that passes the gates, and generalization must be tested on problems the agent never saw. We posed two optimization problems, structure relaxation and reaction paths by the climbing-image nudged elastic band (CI-NEB)\cite{Henkelman2000}, with forces from GFN-xTB tight-binding methods\cite{GFN1,GFN2}. The metric is the mean number of force evaluations needed to converge a unit (one displaced structure or one reaction). A unit that fails is charged the cost of a run that reaches the cap: 500 calls for relaxation and 7,009 for paths (1,000 steps on seven interior images plus nine initial evaluations). The gate, which runs outside the agent's code, requires convergence on every optimization-set unit the reference converges and rechecks each returned structure or band (Table 1; Methods).

For each problem we varied three things (Table 1; Fig. 1b). The starting point is either an ASE method (library start), which the agent must improve within its family so that a gain improves the method users already run, or a minimal steepest-descent optimizer that the agent may replace entirely (from scratch). The ASE methods are BFGS\cite{NocedalWright2006}, FIRE\cite{Bitzek2006}, L-BFGS\cite{LiuNocedal1989} (ASE's LBFGS) and BFGS with a line search (BFGSLineSearch) for relaxation, and CI-NEB driven by FIRE or ODE12r\cite{Makri2019} for paths. The search protocol is either single-phase, the loop described above, or two-phase, which first tunes parameters alone (the floor) and then retunes every new mechanism before judging it. Both keep only improvements, but a single-phase loop judges a new mechanism with constants tuned for the previous one and can discard one that would win once tuned. Finally, informed runs could read the record of every non-informed run on the same problem, from both starting points; all other runs saw only their own. Opus 5 ran both protocols from both starting points, three runs of 500 iterations each. Claude Sonnet 5 and Claude Fable 5.1 repeated the single-phase runs from scratch, and Opus 5 and Fable 5.1 made the informed runs (100 iterations). We call the code at the end of a run its final.

\begin{table}[tbp]
\caption{\textbf{Tasks, starting points, metrics, gates and data sets.}}
\label{tab:1}
{\scriptsize
\begin{tabularx}{\linewidth}{>{\hsize=0.500\hsize\raggedright\arraybackslash}X>{\hsize=1.200\hsize\raggedright\arraybackslash}X>{\hsize=1.150\hsize\raggedright\arraybackslash}X>{\hsize=1.150\hsize\raggedright\arraybackslash}X}
\toprule
\textbf{\hspace{0pt}} & \textbf{\hspace{0pt}implementation: kernels} & \textbf{\hspace{0pt}method: structure relaxation} & \textbf{\hspace{0pt}method: reaction paths} \\ \midrule
\hspace{0pt}\textbf{tasks} & \hspace{0pt}six kernels: effective-medium theory (EMT)\cite{Jacobsen1996} and Lennard-Jones energy and forces (emt, lj), neighbor lists (neighborlist), minimum-image distances (geometry), bond, angle and radial-distribution-function analysis (analysis), the image-dependent pair potential (IDPP)\cite{Smidstrup2014} (idpp) & \hspace{0pt}BFGS\cite{NocedalWright2006}, FIRE\cite{Bitzek2006}, LBFGS\cite{LiuNocedal1989}, BFGS\-Line\-Search (library start); a relaxer from scratch & \hspace{0pt}CI-NEB\cite{Henkelman2000} driven by FIRE or ODE12r\cite{Makri2019} (library start); a path optimizer from scratch \\
\addlinespace[4pt]
\hspace{0pt}\textbf{starting point} & \hspace{0pt}ASE as released & \hspace{0pt}ASE's method, kept within its family, or a minimal steepest descent & \hspace{0pt}ASE's method, kept within its family, or a minimal steepest-descent CI-NEB \\
\addlinespace[4pt]
\hspace{0pt}\textbf{metric} & \hspace{0pt}wall time, median over perturbed fresh inputs & \hspace{0pt}force calls per unit to a largest atomic force of 0.01 eV/\AA{}; 500 on failure & \hspace{0pt}force calls per reaction to a largest image force of 0.05 eV/\AA{}; 7,009 on failure \\
\addlinespace[4pt]
\hspace{0pt}\textbf{in-loop gate} & \hspace{0pt}outputs match stock ASE ($10^{-10}$ absolute, $10^{-9}$ relative); a subset of the module's ASE tests passes; adversarial review & \hspace{0pt}converges every unit the reference converges; force rechecked & \hspace{0pt}converges every reaction the reference converges; band rechecked (force, collapse, endpoints, segment share) \\
\addlinespace[4pt]
\hspace{0pt}\textbf{optimization set} & \hspace{0pt}8--14 systems per kernel & \hspace{0pt}17 systems $\times$ 3 displacements = 51 units (GFN1-xTB\cite{GFN1}) & \hspace{0pt}20 Birkholz reactions\cite{Birkholz2015} (GFN2-xTB\cite{GFN2}) \\
\addlinespace[4pt]
\hspace{0pt}\textbf{held-out set} & \hspace{0pt}none for speed; ASE's test corpus (481 tests) for correctness & \hspace{0pt}10 new systems $\times$ 3 displacements = 30 units (GFN1-xTB) & \hspace{0pt}19 Baker reactions\cite{Baker1996} (GFN2-xTB) \\
\bottomrule
\end{tabularx}
\normalsize}
\par\smallskip{\scriptsize\setlength{\parskip}{0pt}The agent sees only the optimization set; the value of record is the head of the run's branch, evaluated after the run ends. The held-out systems and reactions were fixed before the first run. They omit five of the 24 Baker reactions distributed by Marks and Gomes\cite{MarksGomes2025}, which duplicate Birkholz reactions under other names. Relaxation units start from structures with Gaussian random displacements (0.20 \AA{} on the held-out set; Methods).\par}
\end{table}

\begin{figure}[tbp]
\centering
\includegraphics[width=1.0\linewidth]{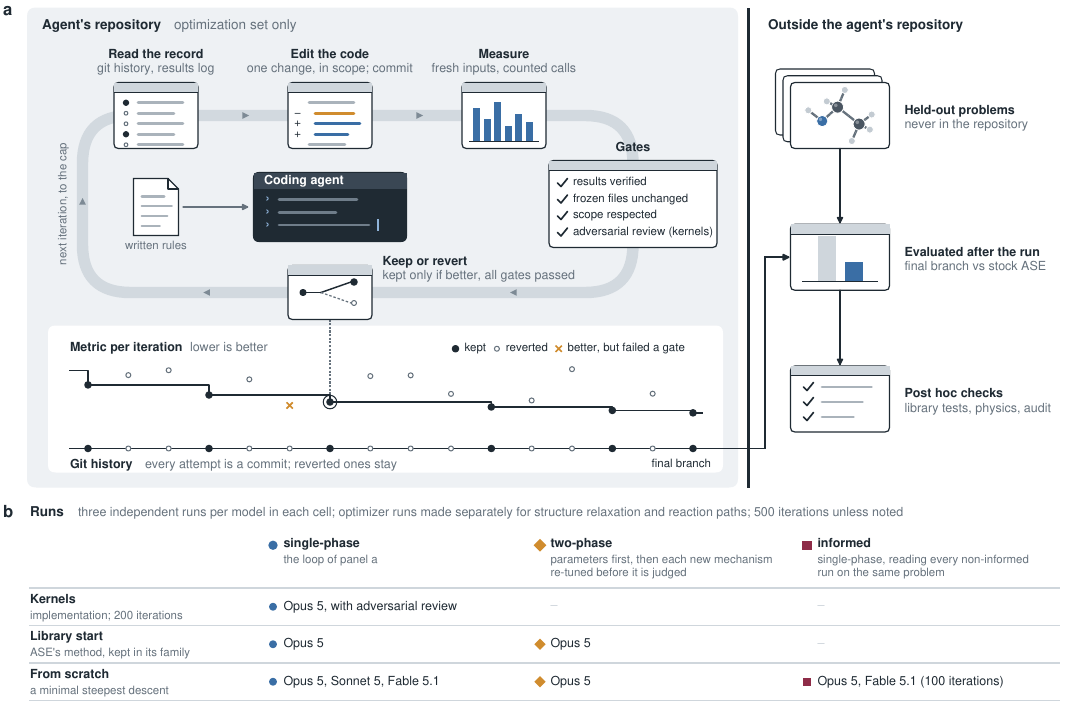}
\caption{\textbf{The AIOA loop, its safeguards and the runs.} \textbf{a}, The loop (left) and what lies outside the agent's repository (right). A change is kept only if the metric improves and every gate passes; reverted commits stay in the history the agent reads (bottom, schematic). The safeguards act on the agent (written rules), on the measurement (fresh, perturbed inputs for wall time; every calculator call counted) and through the gates (for the kernels, also an adversarial reviewer). Each optimizer final is evaluated on the held-out problems after the run, and post hoc checks follow (Methods). \textbf{b}, The runs, by starting point and search protocol; optimizer runs were made separately for structure relaxation and reaction paths. The marker shapes recur in Figs. 3 and 5 and Supplementary Fig. 1.}
\label{fig:1}
\end{figure}

\needspace{10\baselineskip}\subsection*{Faster kernels with unchanged results}
In Python with NumPy and SciPy, the three runs made the six kernels 23.9-, 24.1- and 21.8-fold faster than stock ASE (geometric means over the kernels; 23.2-fold across runs), measured together on one node and one core (Fig. 2). Per-kernel gains ranged from 12-fold (minimum-image geometry) to 99-fold (IDPP\cite{Smidstrup2014}, the pair potential whose minimization builds initial bands). The runs reached the same aggregate with different code. Among runs that kept a kernel in Python, the gains differed by up to 2.3-fold (Lennard-Jones 15--35-fold), and no run was best on every kernel. For each kernel the runs found the same main change, such as array operations over atom pairs in place of per-atom Python loops, and differed in what followed (Supplementary Table 1).

The instructions neither allowed nor forbade rewriting a kernel in another language, so the choice fell to the agents. Run 1 compiled C kernels for three of the six tasks (two more call its compiled pair search), run 3 for one and run 2 for none. The delivered aggregates of runs 1, 2 and 3 were 62.4-, 24.1- and 29.0-fold. We report the gains without the agents' C as representative, since they compare like with like across runs.

\begin{figure}[tbp]
\centering
\includegraphics[width=1.0\linewidth]{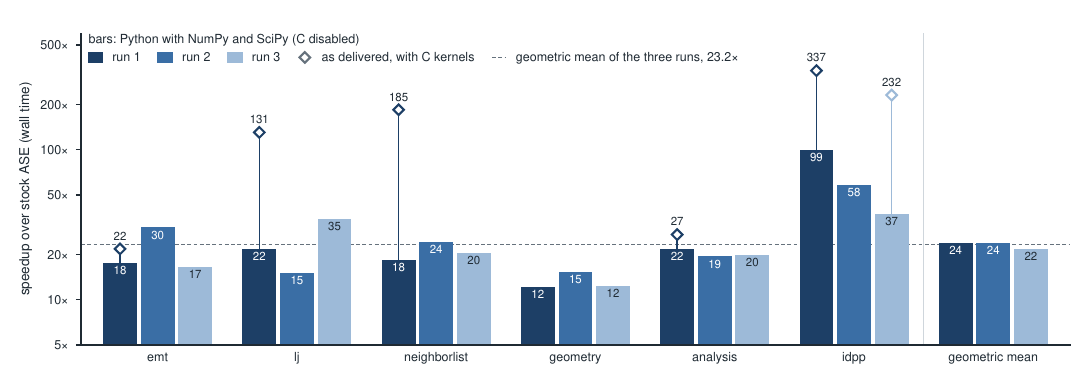}
\caption{\textbf{Faster ASE kernels, reached by different code in each run.} Wall-time speedup over stock ASE for each kernel and their geometric mean, all trees measured on one node and one core. Bars, each run's merged tree in Python with NumPy and SciPy, with the agents' C disabled. Diamonds, the tree as delivered, for the kernels that use the agents' C. Run 1 compiled lj, neighborlist and idpp, and its emt and analysis kernels call the compiled pair search; run 3 compiled idpp. Dashed line, the geometric mean of the three runs' bars (23.2-fold).}
\label{fig:2}
\end{figure}

\needspace{10\baselineskip}\subsection*{Fewer force evaluations on held-out problems}
On the blind held-out problems, the Opus 5 finals of both protocols needed 1.81-fold fewer force evaluations than each task's ASE reference (geometric mean over tasks): 1.63-fold for structure relaxation (five tasks) and 2.15-fold for reaction paths (three tasks; Fig. 3; Supplementary Table 2). The reference is the stock method the task started from or, from scratch, the stock method with the fewest force calls on the optimization set. No Opus 5 path final of either protocol converged fewer reactions than its reference, and four converged all 19. Two relaxation finals each left one of the 30 units unconverged where their reference converged all.

Path means need care. Stock CI-NEB with FIRE and with ODE12r each failed three of the 19 reactions, and the 7,009-call penalty for these failures makes up half of their means. Without penalty, on the reactions that both a final and its reference converge, every Opus 5 path final of either protocol still needed 1.5--4.5-fold fewer calls (Supplementary Fig. 1).  Unless stated otherwise, values use the pre-specified metric.

With the same model, the from-scratch finals did not surpass the best library-start finals (means of six runs: 30.6 force calls per relaxation unit against 26.5 for BFGS; 1,129 per reaction against 903 for CI-NEB with ODE12r; Fig. 3). Freedom to replace the method did not by itself produce larger gains. The largest step came from the informed runs.

Starting from scratch for 100 iterations but free to copy from the records of all non-informed runs on the same problem, the three Fable 5.1 informed runs needed 15.0--16.4 force calls per held-out relaxation unit and 128--481 per reaction, against 50.5 and 2,236 for their references, L-BFGS and CI-NEB with ODE12r (Fig. 3): 3.07--3.37- and 4.6--17.5-fold fewer. They converged 18 or 19 of the 19 reactions and, on those that both converged, needed 7.5--12.0-fold fewer calls. Against BFGS with a line search, the stock relaxer with the fewest held-out calls, they needed 2.34--2.57-fold fewer. The three Opus 5 informed runs reached 2.94--2.99-fold on relaxation and 2.39--14.3-fold on paths.

Two checks address simpler explanations. First, tuning parameters alone does not reach these results. The best parameter floors (stock methods with parameters tuned on the optimization set; Methods) need 2.0--2.3 times the informed relaxers' force calls and 1.9--7.3 times those of the informed path optimizers, except Opus 5 run 2, which only matches the best floor (935.8 against 937.8 calls).

Second, the gains do not come from stopping early. Every optimizer was asked to converge bands to 0.05 eV/\AA{} but measures this with its own tangent, and the held-out gate accepts forces up to 0.15 eV/\AA{} (Methods), so a final could stop early and still pass. Re-evaluated the same way for every optimizer, from fresh calculator forces and with the gate's tangent, the informed finals' bands stopped at forces no higher than stock CI-NEB's (both forces at most 0.060 eV/\AA{}, against up to 0.067 for stock). Looser stopping does occur among the library-start finals, whose 90th-percentile perpendicular force is 0.067 eV/\AA{}, against 0.050 for the informed finals (Supplementary Table 3).

\begin{figure}[tbp]
\centering
\includegraphics[width=1.0\linewidth]{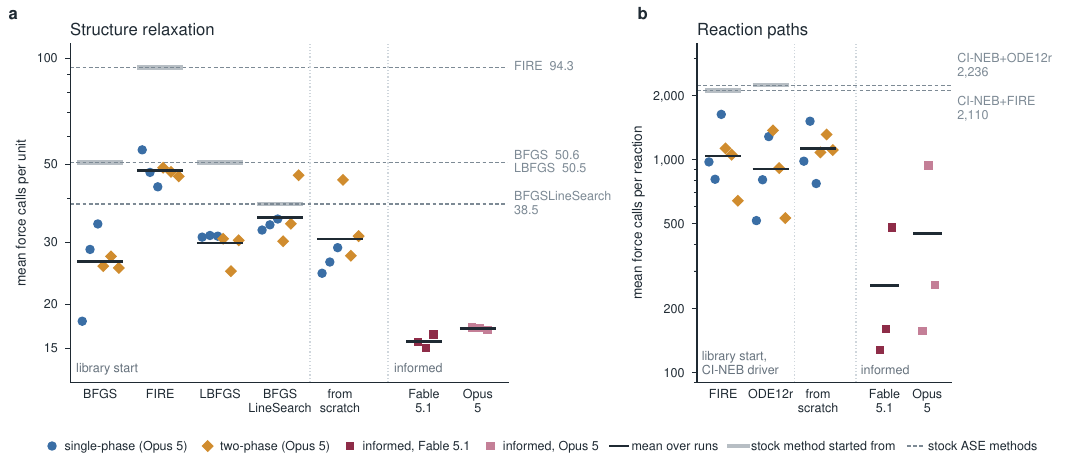}
\caption{\textbf{Fewer force evaluations on blind held-out problems.} Each marker is one run's mean force calls per held-out unit; each unconverged unit is charged 500 (relaxation) or 7,009 (paths) calls, and path bands must pass the held-out gate (Methods). \textbf{a}, Structure relaxation, 30 units. \textbf{b}, Reaction paths, 19 reactions; library-start tasks start from CI-NEB driven by FIRE or ODE12r. Informed runs also start from scratch but may copy archived code; Sonnet 5 and Fable 5.1 single-phase runs from scratch are in Fig. 5. Dashed lines, stock ASE methods at default parameters. Speedups of runs from scratch and informed runs are quoted against L-BFGS (ASE's LBFGS) and CI-NEB with ODE12r, the stock methods with the fewest force calls on the optimization set. Penalty-free speedups are in Supplementary Fig. 1.}
\label{fig:3}
\end{figure}

\needspace{10\baselineskip}\subsection*{The safeguards in action}
The gates rejected many candidates that improved the metric but were wrong. Counting those discarded because a gate, or a check of a gate's behavior, exposed a defect (Methods), we found at least 53 for the kernels. The adversarial reviewer caught 51, including a k-d-tree pair search, 7.75 times faster, whose output differed from stock ASE's on rank-deficient periodic cells. After the campaign, all three merged kernel trees passed 480 of the 481 kernel-relevant tests in ASE's own suite. The one failure, ASE's finite-difference force test, is sensitive to round-off and also fails in stock ASE with another random seed.

For the optimizers we found at least 42. Thirty-three failed the path verification, with degenerate or unresolved bands or forces above the tolerance. One of them delayed the climbing image and halved the force calls but left a single segment carrying 97.6\,\% of the path. The blind held-out set then caught five optimizer finals that every in-loop gate had accepted, refusing six of their bands. Two had loosened convergence for the climbing image alone, leaving the force on the highest image of each refused band at 3.3--3.9 times the 0.05 eV/\AA{} convergence threshold; one of them, at 438 calls per reaction, would have been the best Opus 5 path result outside the informed runs, and its value of record is 1,109. In the other three, informed Fable 5.1 run 3 among them, one band each had a single segment carrying 64--67\,\% of the path.

A mechanical check of the full history of every kernel and optimizer run (the 80 optimizer runs, including the eight parameter floors, and the three kernel runs) found no change to a frozen file or outside scope, and source audits of the strongest runs found no rule violations.

\needspace{10\baselineskip}\subsection*{What the agents built}
How much of the gain is algorithmic depends on the task. To separate tuning from algorithmic change, a parameter-only run on each optimizer task (the floor; one run, at most 100 iterations) tuned the numeric parameters of the starting method without changing its code. We compare the floor with the finals (Fig. 4a; Supplementary Table 2). On five of the six library-start tasks, tuning alone gets most of the way. On held-out problems the floor reaches 77--107\,\% of the finals' speedup. On the optimization set, algorithmic change accounts for only 7--30\,\% of two-phase run 1's improvement on these tasks. On CI-NEB with FIRE and from scratch the picture reverses. Tuning alone made held-out results worse than the starting point, and algorithmic change accounts for 86--96\,\% of the improvement on the optimization set.

Agents reassembled published methods into new optimizers. All six single-phase Opus 5 and Fable 5.1 relaxers from scratch independently derive a model of the curvature from the geometry, each coded anew rather than imported. Three use an exponential pair-stiffness model of Packwood et al.'s form\cite{Packwood2016}, one the stretch terms of Lindh's model Hessian\cite{Lindh1995} and two valence force fields with bends and torsions. Most use these models in published ways (Methods). The three Sonnet 5 relaxers instead rebuilt textbook quasi-Newton methods from an isotropic starting matrix (Fig. 4b; Methods). Having such a model is not enough. ASE's preconditioned L-BFGS, which uses the exponential form, needs 78.4 force calls per held-out unit it converges, against 15.1--17.3 for the informed relaxers on those units (post hoc and untuned, non-periodic units placed in a box; Supplementary Table 4). The informed Fable 5.1 path optimizers, the best configuration, were assembled from the record: all three adopted the band driver of the archived final with the lowest optimization-set score, and the coarse-to-fine schedule and spline climbing-image tangent of another (Methods). We found no published match for any complete optimizer or for two of their components.

Of the two components without a published match, one is in a relaxer (Opus 5, single-phase, run 1): its L-BFGS initial inverse Hessian blends a bond-graph preconditioner with the standard scaled identity, $w\,\mu^{-1}P^{-1} + (1-w)\,\gamma I$, and refits the weight $w$ at every step by least squares over the five most recent curvature pairs (Methods). With the weight fixed post hoc, the held-out mean rose from 24.5 to 28.5 force calls for the preconditioner alone and to 34.1 for the scaled identity alone. At the constants tuned with it, the blend beats either part. The other is in path optimization: two Fable 5.1 runs from scratch preconditioned each image with a model of Lindh's form, which all three informed Fable 5.1 runs extended to the full model Hessian, a combination of two published ideas\cite{Mones2018,Makri2019} that we did not find in the literature; we did not isolate its share of their gain.

\begin{figure}[tbp]
\centering
\includegraphics[width=1.0\linewidth]{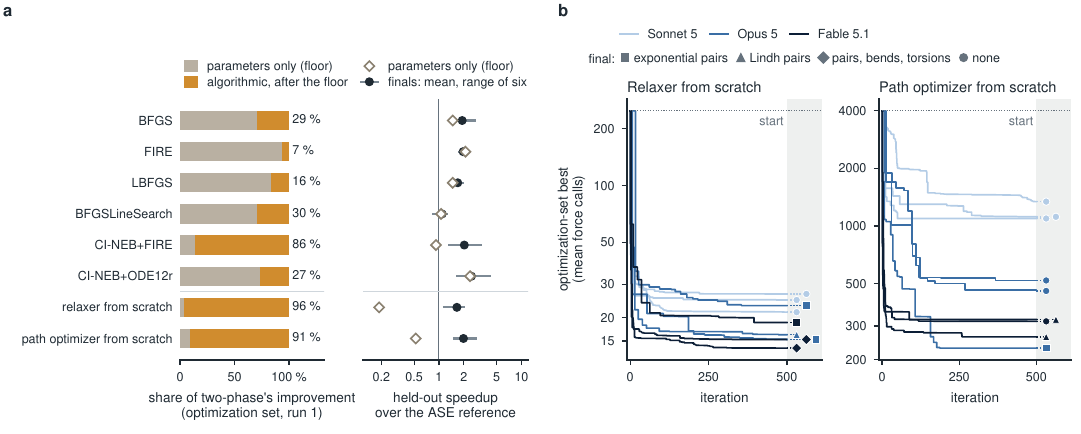}
\caption{\textbf{What the agents built.} \textbf{a}, Left, share of two-phase run 1's optimization-set improvement reached by tuning parameters alone (the floor) and by algorithmic change after it; numbers, the algorithmic share (Methods). Right, held-out speedup over the reference of Fig. 3 for the floor (one run per task) and the finals (mean of six single-phase and two-phase runs, line showing their range; Supplementary Table 2). \textbf{b}, Optimization-set best against iteration for the nine single-phase runs from scratch of each task (three per model); horizontal dotted line, the starting optimizer. Beyond iteration 500 (shaded), dotted extensions lead to each final's marker, offset sideways where markers would overlap. Marker shape, the curvature model read from the final's code (Methods).}
\label{fig:4}
\end{figure}

\needspace{10\baselineskip}\subsection*{Variance across runs and models}
Results vary from run to run, so we report runs rather than single best results. Within an optimizer task, the best of six runs needs 1.3--2.7 times fewer force calls than the worst (Fig. 3). Reporting the best run per task would give 2.46-fold across tasks instead of the 1.81-fold of the mean of six, a gain from selection, not from the method. The two protocols did not differ detectably: the ratio of single-phase to two-phase held-out means is 0.98 (run-level bootstrap 95\,\% interval 0.87--1.10). The logs suggest one reason. Single-phase runs also retuned constants after adding a mechanism, in the next iterations or within the same change, so the protocols differed less in practice than in design.

Differences between models were larger. On both problems, every single-phase run from scratch of Sonnet 5, the lowest-priced of the three models (Methods), did worse on held-out problems than every such run of Opus 5 and of Fable 5.1 (Fig. 5), the strongest separation three runs against three can show (p = 0.10 for each pairwise comparison; Methods). Sonnet 5 still needed 1.28--1.42-fold fewer force calls than L-BFGS for relaxation, but on paths it stayed near CI-NEB with ODE12r (0.94--1.40-fold), and one run converged 14 of 19 reactions, two fewer than stock (Supplementary Table 2). On paths the gap already shows on the optimization set, where Sonnet 5's finals needed 1,094--1,340 calls per reaction against at most 519 for the other models, so the loop's own score flags the weaker result before any held-out evaluation. Opus 5 and Fable 5.1 overlapped from scratch (Fig. 5).

\begin{figure}[tbp]
\centering
\includegraphics[width=0.54\linewidth]{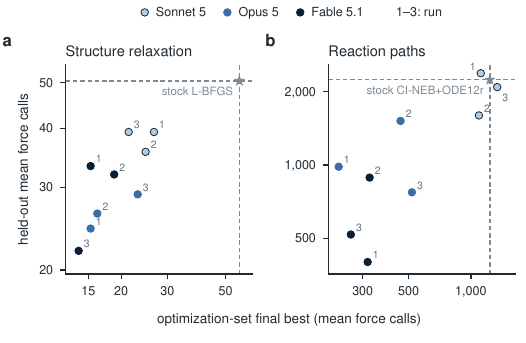}
\caption{\textbf{Three models from scratch.} Held-out value of record against the final optimization-set best for the single-phase runs from scratch (three runs per model, 500 iterations), in mean force calls per unit: \textbf{a}, structure relaxation (held-out units at 0.20 \AA{} displacements); \textbf{b}, reaction paths. Star and dashed lines, the stock reference, the stock method with the fewest force calls on the optimization set (L-BFGS; CI-NEB with ODE12r). Log axes. Spearman $\rho$ between the two scores: 0.75 (\textbf{a}) and 0.73 (\textbf{b}) over all nine runs, 0.49 and 0.14 over the six Opus 5 and Fable 5.1 runs.}
\label{fig:5}
\end{figure}

\section*{Discussion}
Working autonomously, agents made the kernels of a widely used scientific library faster and its optimizers cheaper in force evaluations on unseen problems, with gains that recurred across independent runs (Figs. 2 and 3). Force evaluations dominate the cost of relaxations and band calculations with first-principles calculators, which take minutes to hours per evaluation, so gains that carried over to these calculators would save compute directly. With cheap calculators such as GFN1-xTB, the optimizer's own work also counts (15--43\,\% of the informed relaxers' wall time on the molecules, though they still finished 1.9--3.9-fold sooner than L-BFGS; Supplementary Table 5), and an objective may need to weigh it. The kernel gains measure the room left in ASE's Python code (four of the six kernels loop over atoms; Methods), not the distance to the fastest possible implementation. Compiled code goes further (Fig. 2). Much scientific code is written in high-level languages, and agents can recover such room on their own under checks they cannot alter.

With three runs per configuration, comparisons are descriptive, and differences below about 10--15\,\% are invisible. Each change is kept or reverted on a small optimization set, whose score tracked the held-out results only loosely for the six Opus 5 and Fable 5.1 runs of Fig. 5 (Spearman $\rho$ 0.49 and 0.14). Every relaxation unit starts near an equilibrium structure, so the relaxers' gains are shown within one basin. On the drug-like molecules of concurrent work\cite{AutoSella}, started far from their minima, the informed relaxers needed more force evaluations than Sella and failed to converge more often. On our held-out molecules, they needed as many as AutoSella-A, the best of that work's optimizers there, or fewer (Supplementary Table 6).

Each gain comes with a record of how it was obtained and checked, and three design elements that produce this record transfer to other domains. The first is a set of safeguards independent of the optimizing agent: a numeric check on fixed systems is necessary but not sufficient, and an adversarial reviewer that had to execute its counterexample caught most kernel defects. The second is a blind held-out set, kept separate by placement, rules and audit rather than access control. The third is reporting across runs, as one run could have supported a different headline. The record has a further use: the informed runs that read it copied and recombined what had worked, and 10 of 12 beat every archived final (Methods). A record of agents' successes and failures may thus be a resource for designing better algorithms, not only an audit trail.

\section*{Methods}
\needspace{10\baselineskip}\subsection*{The AIOA loop}
AIOA runs one agent per task in a git repository that holds a local copy of the library (\texttt{ase\_local/}), the benchmark and gate scripts and a results log. Each run's instructions fix the goal, the scope (the files the agent may change), the metric, the gate command and the iteration cap. One iteration is one candidate evaluated once on the whole optimization set.

The search is the keep-or-discard loop of autoresearch\cite{autoresearch}, with gates added to the keep test. In each iteration the agent reads the results log and the git history, including discarded attempts, makes one change inside the scope and commits it before measuring. It then runs the benchmark and the gates and keeps the commit if the metric improves and every gate passes, or reverts it with \texttt{git revert}, so that the failed attempt stays in the history. Each iteration appends a row to \texttt{results.tsv} with the metric, the incumbent best, the action, a category (rows tagged as cheating are discarded) and the cumulative count of unit-runs (Compute accounting).

\needspace{10\baselineskip}\subsection*{Tasks and data sets}
Table 1 lists the tasks. All runs use a local copy of ASE 3.28.0 with Python 3.11, NumPy 2.4.3 and SciPy 1.17.1. The six kernels, each timed on 8--14 benchmark systems, cover distinct kinds of computation in ASE: energy and forces from two classical potentials (EMT\cite{Jacobsen1996} and Lennard-Jones), the neighbor-list function that several calculators and the preconditioned optimizers build on, minimum-image distances under periodic boundary conditions, structural analysis (bonds, angles and radial distribution functions) and the image-dependent pair potential (IDPP) that builds initial bands for reaction paths. Their stock implementations leave room for speedup without changing results: four (EMT, Lennard-Jones, IDPP and the analysis) loop over atoms in Python, and the other two, though largely vectorized, still loop in Python over neighboring bins or test up to 27 periodic translations for every displacement in general cells.

Relaxation units are built from 17 optimization-set systems (bulk metals and an alloy, a Cu$_{55}$ cluster, diamond, small molecules, an organometallic complex, $\pi$-stacked and hydrogen-bonded dimers, and solid argon), each a reference structure (an experimental lattice constant, or a database or hand-built geometry) given Gaussian random displacements (standard deviation 0.05--0.20 \AA{} per coordinate, set per system) for three displacement seeds (42--44), giving 51 units. All units use GFN1-xTB\cite{GFN1} (xtb 6.7.1\cite{XTB2021} through xtb-python 22.1). A unit converges at fmax < 0.01 eV/\AA{}; a unit that fails or exceeds 500 force calls is charged 500. The held-out set is 10 systems with new elements, molecules and bonding motifs (Pd and Ag bulk, an Au$_{55}$ cluster, methanol, furan, ammonia, methylsilane, and formic-acid, phenol and pyrazine dimers) with the same seeds (30 units). Its units are displaced with a standard deviation of 0.20 \AA{} for every system, the largest amplitude of the optimization set.

The reaction-path tasks use 20 reactions of the Birkholz--Schlegel set\cite{Birkholz2015} for optimization and the Baker set\cite{Baker1996} for held-out evaluation, both as distributed with the freezing-string-method code of Marks and Gomes\cite{MarksGomes2025} (24 of Baker's 25 reactions). Both use GFN2-xTB\cite{GFN2}. Five Baker reactions are Birkholz reactions under other names (endpoints identical within 0.012 \AA{}) and were excluded on 15 August 2026, two weeks before the first run, leaving 19. Each reaction starts from a nine-image band built once (endpoints relaxed with GFN2-xTB, interior images interpolated by IDPP) and reused unchanged by every evaluation, so that the interpolation cannot be re-optimized. Charge and spin multiplicity come with each reaction's structures, and xTB otherwise runs at the xtb-python defaults (the path gates allow up to 1,000 SCF iterations). A band converges when its forces fall below 0.05 eV/\AA{}; a reaction not converged within 1,000 optimizer steps is charged 7,009 calls, the cost of a cap-bound run (1,000 steps on seven interior images plus nine initial evaluations).

Because every candidate is measured on the whole optimization set, up to 500 times per run, we kept the optimization sets small. The held-out sets are different problems (other systems, another reaction set) and are evaluated only after a run ends. The optimization set is open to the agent, as training data would be: its structures, references and per-system results are in the agent's repository. Every unit runs as its own single-core job (one thread for every numerical library), because step counts on long trajectories are sensitive to the summation order of multi-threaded calculators.

\needspace{10\baselineskip}\subsection*{Safeguards by construction}
\textbf{Rules.} A project instruction file that Claude Code loads into every session (\texttt{CLAUDE.md}) defines cheating, and the agent applies the definition to its own candidates. The file forbids memoizing outputs or input-keyed intermediates, exploiting the benchmark's call pattern, editing benchmark, gate or reference files, and, for force-call metrics, obtaining forces or energies in any way without a counted calculator call (for example by overriding \texttt{check\_state}, wrapping the calculator or estimating forces by finite differences).

\textbf{Benchmark design.} Each wall-time benchmark reports the median of 20 timed runs after two warm-ups. Every timed run builds fresh copies of the inputs with a $10^{-10}$ \AA{} random perturbation and its own seed, so caches keyed on object identity or input bytes miss. Force-call benchmarks count every call to the calculator's \texttt{calculate()} method, which ASE runs at each new geometry; repeated queries at one geometry, served from ASE's own cache, are not counted.

\textbf{Frozen files and scope.} At the start of a run the loop records SHA-256 hashes of every reference file, vendored unit test and Python file under \texttt{benchmarks/} (the force counter included) and of the pytest configuration. After every iteration it re-checks them and asserts that nothing outside the scope changed; a mismatch reverts the iteration, cleans untracked files and logs the iteration as cheating.

\needspace{10\baselineskip}\subsection*{Gates}
\textbf{Kernel verification.} On each benchmark system the modified kernel's outputs (for example energy, forces and, for three-dimensional cells, stress) must match frozen stock-ASE references to an absolute tolerance of $10^{-10}$ and a relative tolerance of $10^{-9}$, which admits floating-point reordering and rejects a single-precision downgrade. The gate also runs a compressed subset of ASE's unit tests for the module, vendored into the benchmark and importing the local copy.

\textbf{Adversarial review (kernels).} Every candidate that passes verification and would be kept is examined by a separate agent (Opus 5; in run 1 also Claude Sonnet 5.5) that sees only the diff and the in-scope source, not the optimizer's reasoning or results. It searches for inputs on which the change diverges from stock ASE (boundary, degenerate, empty or unusual inputs) and must execute each counterexample. Only a divergence confirmed by execution discards the candidate. Confirmed counterexamples are queued for the operator, not the agent, to add to the frozen gate; none was added during the three runs.

\textbf{Relaxation verification.} On every optimization-set unit whose reference converges, the optimizer must report convergence within the cap, and an external check on the returned geometry must find fmax at most 1.1 times the threshold. Any force-converged minimum is accepted.

\textbf{Path verification.} The benchmark saves every band, and the gate, which lies outside the agent's scope, re-evaluates each with the raw calculator and its own tangent. On every optimization-set reaction whose reference converges, it requires (1) the true perpendicular force on every interior image and (2) the full force on the highest-energy interior image to be at most 0.15 eV/\AA{}, with finite energies, no atomic clash and no collapsed images; (3) unmoved endpoints; and (4) no single segment carrying more than half of the total path change, with segments measured as Kabsch-aligned root-mean-square displacements between neighboring images. The clash, collapse and segment checks also apply to every other band the gate finds converged, and a reaction on which the gate's own force evaluation fails is left unchecked, its force calls still counted in the metric. The tolerance is three times the convergence threshold because each optimizer measures convergence with its own tangent and preconditioner, so a band converged by its own measure can read higher on the gate's. The tolerance still rejects an unrelaxed band (about 1 eV/\AA{}) or a climbing image stranded below the saddle. Saddle position, barrier height and path geometry are not checked: the nudged elastic band has no unique solution, and a different valid saddle, including a lower one, is a correct answer.

\needspace{10\baselineskip}\subsection*{Protocols and runs}
All run agents used Claude Code with high reasoning effort: Claude Fable 5.1 (\texttt{claude-fable-5-1}), Claude Opus 5 (\texttt{claude-opus-5}) and Claude Sonnet 5 (\texttt{claude-sonnet-5}), in descending order of Anthropic's list price\cite{AnthropicModels}. Every run had its own repository, started from a clean copy of the framework, and its instructions forbade reading anything outside it.

In each of the three kernel runs, a coordinating agent ran the six kernel agents (Opus 5) in parallel, each in a single-phase loop with adversarial review for up to 200 iterations, and then merged the six branches into one tree. The instructions were identical across runs.

The library-start instructions keep each method in its family: BFGS stays quasi-Newton (an approximate Hessian $B$ and a step $-B^{-1}g$, with $g$ the gradient); FIRE keeps its molecular-dynamics nature (velocity state, force-only input); LBFGS keeps a finite history of curvature pairs without a stored Hessian; BFGSLineSearch may change only its line search; and the band tasks may change the NEB algorithm and its driver but keep a chain of images joined by springs, ODE12r with its error-controlled adaptive solver. The from-scratch tasks start from a minimal steepest-descent optimizer (an 81-line relaxer or a 157-line CI-NEB) that the agent may replace entirely.

A single-phase run is the loop above for 500 iterations. A parameter-only run (one per task) changes numeric values only, never control flow, and stops at 100 iterations or once a full pass over its parameters brings no improvement; its best values are the floor $F$. A two-phase run starts from $F$: each new mechanism enters with default constants and is re-tuned, for at most 20 iterations, before it is judged, within 500 iterations in total (one run overran to 561). Both protocols keep a change only if it improves the metric, and neither stops on a plateau: each run is meant to spend its cap. Opus 5, Sonnet 5 and Fable 5.1 each made three single-phase runs from scratch with identical instructions, compared at equal iteration counts; every other non-informed optimizer run used Opus 5.

The informed runs follow the instructions of the single-phase runs from scratch with two changes: the agent may read a frozen, explicitly listed archive of every non-informed run on the same problem, held as branches in its repository (41 relaxation and 27 path branches, the library-start runs and parameter floors included), with their code, logs and history, and the cap is 100 iterations. Anything that would reveal the held-out set remains forbidden. Opus 5 and Fable 5.1 each made three informed runs on each problem. Every informed relaxer and four of the six informed path optimizers needed fewer held-out force calls than any archive member (relaxers by 3.9--16\,\%, path optimizers by 35--68\,\%); of the other two, Opus 5's had two calculator crashes and Fable 5.1's a band refused by the path checks, both scored as failures. Without a continuation control, reading the record cannot be separated from continuing from a strong earlier final, and the informed runs stand on the compute of the archive they read.

No operator intervention gave method guidance. The operator resumed optimizer runs that stopped before their cap by quoting the rule that a run spends its whole budget, and sent the runs of one model brief operational notes, for example on cluster throughput; the Opus 5 optimizer runs received none. In the kernel runs the operator resumed run 3's coordinator twice after usage pauses and accepted three tasks it closed before the cap (at iterations 150--179). The coordinators, themselves agents, also steered their kernel agents: run 1's pushed the IDPP kernel toward C and narrowed the reviewer's brief to well-posed inputs, run 2's declined one proposal to compile, and run 3's relayed ideas between tasks and restricted candidates.

\needspace{10\baselineskip}\subsection*{Kernel measurement}
The values of record come from one joint re-measurement after all runs ended. Every tree (stock, each task head, each merged tree, and the merged trees with native code disabled) was exported with \texttt{git archive} (no bytecode or compiled objects) and timed on one exclusive node (Intel Xeon Gold 5118, one core) in five interleaved rounds after a warm-up run, with the median taken per tree. Speedup is the stock median divided by the tree median; aggregates are geometric means over the six kernels. ``Native-off'' hides every compiler and switches off the agents' C kernels, so each falls back to its Python path, which uses NumPy and SciPy, both dependencies of ASE. Stock ASE already calls SciPy, and all three runs added calls to its compiled routines (k-d tree, sparse matrices, LAPACK). Once an agent moved a kernel to C it stopped tuning the Python fallback, so the native-off values of the two compiling runs are lower bounds.  A second node of the same model gives 5--12\,\% higher speedups.

\needspace{10\baselineskip}\subsection*{Held-out evaluation of the optimizers}
The value of record for each run is its branch head, evaluated on the held-out units after the run ended, with the same caps and penalties and, on paths, the path gate. The metric, caps, penalties and gates were fixed in the frozen benchmark and evaluation code before the runs began.

Library-start finals are compared with the stock ASE method for their task, run at ASE's default parameters (for paths with the improved tangent and the climbing image on from the start). From-scratch and informed finals are compared with the stock method that needs the fewest force calls on the optimization set, a choice that uses no held-out data: for relaxation L-BFGS (56.4 calls per unit, against 62.4--88.0 for BFGS, BFGSLineSearch and FIRE; 50.5 on the held-out units), reported beside BFGSLineSearch, the stock method with the fewest held-out calls (38.5); for paths CI-NEB driven by ODE12r (1,235 calls per reaction, against 2,769 with FIRE; on the held-out reactions 2,236 and 2,110). The primary statistic is the ratio of the reference's mean force calls per held-out unit to the final's, with penalties. Where the reference or a final fails units (every path task, and two relaxation finals), we also report the ratio of means over the units both converge, without penalty (Supplementary Fig. 1).

\needspace{10\baselineskip}\subsection*{Statistics}
With three runs per configuration the statistics are descriptive. For the optimizers, task aggregates are means over runs and cross-task aggregates are geometric means; the kernels' cross-run figure is the geometric mean of the three runs' geometric means. The protocol comparison bootstraps runs within each task for a 95\,\% interval on the aggregate ratio of single-phase to two-phase means. The model ordering is reported as complete separation of Sonnet 5's three runs from the three runs of each other model on the held-out problems (two-sided permutation p = 0.10 for each comparison). Rank correlations are Spearman's $\rho$ with average ranks for ties. The algorithmic share of an optimizer task is $(F - M)/(B_0 - M)$, where $B_0$, $F$ and $M$ are the optimization-set scores of the stock method or starting optimizer, the floor and the two-phase run-1 final.

\needspace{10\baselineskip}\subsection*{Integrity audits and physics checks}
\textbf{Integrity.} Over the full histories of all 80 optimizer runs and the three kernel runs we checked mechanically that no commit touched the frozen benchmark and data trees and that no change left the scope (all clean). We also audited the sources of the strongest kernel and optimizer runs, with their session transcripts where these were kept. Two rules have no mechanical enforcement and rest on these audits: no force or energy may be estimated without a counted call, and nothing outside the run's repository may be read. The held-out material was kept outside every agent repository but on a shared filesystem readable by the agents' user account. The audited transcripts show no read of held-out or orchestrator material.

\textbf{Gate catches.} A candidate counts if it improved the metric and was discarded because a gate, or a check of the gate's behavior, exposed a defect. Defects found by the agent's own checks or repaired in kept code, and failing points inside a parameter sweep, are not counted. We read every discarded \texttt{results.tsv} row that improved on the incumbent and every row recording a review rejection, a failed verification or another correctness failure. This gave 53 catches for the kernels and 42 for the optimizers, lower bounds because rows that record neither an improvement nor a failure are missed.

\textbf{ASE test corpus (kernels).} After the campaign, we ran the kernel-relevant part of ASE's own test suite (481 tests, infrastructure tests excluded) on each merged tree, with the tests importing the local copy.

\textbf{Relaxation physics.} Each held-out relaxation unit of every relaxation final (42 runs) was re-run exactly as in the held-out evaluation, and a fresh calculator gave fmax and energy at the returned geometry; these were compared with stock BFGS and FIRE run without the 500-call cap from the same start. All 1,258 converged units meet the force criterion on the fresh calculator to within its self-consistent-field precision. Of these, 1,239 end no more than 9.4 meV above the lower stock minimum. The other 19 stop higher, as the force criterion allows: 14 on the Au$_{55}$ cluster (up to 76 meV; stock BFGS and FIRE themselves end up to 76 meV apart there) and five on the Ag crystal and the pyrazine dimer (up to 12 meV).

\needspace{10\baselineskip}\subsection*{Post hoc comparisons}
Further stock baselines were run on the held-out set under the conditions of every other held-out evaluation: Sella 2.5.0\cite{Sella} as a minimizer in Cartesian and internal coordinates, ASE's preconditioned relaxer \texttt{PreconLBFGS}\cite{Packwood2016}, and ASE's preconditioned NEB\cite{Makri2019} with the ODE12r driver. The preconditioned NEB ran beside a stock CI-NEB+ODE12r control in the same tree, which reproduced its value of record exactly. Sella and the preconditioned NEB take their best configuration on each unit (Supplementary Table 4). The three optimizers of concurrent work\cite{AutoSella} were run through their own entry point on the 21 molecular held-out units, and seven relaxer finals (the six informed and Opus 5's single-phase run 1 from scratch) were run on that work's benchmark with its harness, through a wrapper that lets the harness's convergence test stop them. AutoSella-F and -A, run the same way, reproduced their published ratios to Sella within 1.2 percentage points (Supplementary Table 6).

\needspace{10\baselineskip}\subsection*{Provenance of the agents' components}
Fig. 4b classes each final by its default code path: a geometry-derived model with bend or torsion terms counts as a valence force field, a pair-only model by the form of its pair stiffness, and anything else as none (one Sonnet 5 path optimizer scales forces by inverse square-root mass only). Of the six single-phase Opus 5 and Fable 5.1 relaxers from scratch, each of which implements its model itself without importing ASE's preconditioner module, three build a geometry-derived pair stiffness $\exp[-A(r/r_0 - 1)]$ ($r$, the interatomic distance; $r_0$, a reference distance; $A$, a decay constant), the form of the preconditioner of Packwood et al.\cite{Packwood2016}, which ASE ships as \texttt{Exp} (one with ASE's own constants); one (Opus 5, run 2) uses the stretch terms of Lindh's model Hessian\cite{Lindh1995}, which ASE does not ship; and two (Fable 5.1, runs 1 and 3) are valence force fields with exponentially weighted stretches, bends and torsions, the form of the model Hessian of Swart and Bickelhaupt\cite{Swart2006}. How the relaxation finals use these models is also largely published: rebuilding the model at the current geometry and keeping only recent curvature, as a production quantum-chemistry code does\cite{OpenMolcas} and a Bayesian quasi-Newton method formalizes\cite{HennigKiefel2013}, or fitting the model's overall scale, as ASE's preconditioner does\cite{Packwood2016}, and updating it by damped BFGS\cite{pysisyphus}. The agents' versions differ in form (an exponential re-blend toward the fresh model, a scale refitted every step).

In the per-step blend of Opus 5's single-phase run-1 relaxer from scratch, the L-BFGS initial inverse Hessian is $H_0 = w\,\mu^{-1}P^{-1} + (1-w)\,\gamma I$, where $\gamma = s^{\mathsf T}y/y^{\mathsf T}y$ and $\mu = s^{\mathsf T}y/s^{\mathsf T}Ps$ come from the newest curvature pair ($s$, the step; $y$, the gradient change), $P$ is rebuilt at the current geometry once an atom has moved more than 0.2 \AA{} since the last build, and $w \in [0, 1]$ minimizes $\sum_k 0.7^k \lVert H_0 y_k - s_k \rVert^2$ over the five most recent pairs ($k = 0$ the newest). For the ablation in Results, the relaxer was re-evaluated post hoc at its record commit with $w$ fixed to 1 or to 0; all 30 held-out units converged in both. The closest relatives of the per-step blend that we found refit a scaled identity added to a structured initial matrix by secant least squares\cite{Mannel2024}, blend a Laplacian against secant data inside the stored pairs\cite{BCQN}, or mix two preconditioners with a fixed weight\cite{Makri2019}; each differs in form.

Lindh's model Hessian and preconditioners built in its form have been used for single-structure optimization and dimer saddle searches\cite{Mones2018}, and per-image preconditioning of the band is published with other matrices\cite{Makri2019}. The archive that the informed Fable 5.1 path runs read held two path optimizers (Fable 5.1, single-phase runs 2 and 3 from scratch) with stretch-only models of Lindh's form, which the informed runs' reports name as their source; the runs then added Lindh's bend and torsion terms. The other path components are published or closely related to published methods: adaptive first-order integration with per-image preconditioning\cite{Makri2019}, selective image relaxation\cite{Lindgren2019} and a coarse-to-fine schedule related to AutoNEB\cite{Kolsbjerg2016}. Mechanisms also spread by copying: all six informed path runs adopted, at their first iteration, the band driver of Opus 5's single-phase run-1 final from scratch, the archive member with the lowest optimization-set score, and all copied the coarse-to-fine schedule, and five the spline climbing-image tangent, of Fable 5.1's run-3 final from scratch. Surrogate-model band methods save force calls by replacing many of them with a fitted model's predictions\cite{Koistinen2017,GarridoTorres2019}. The counted-call rule excludes such predictions, so the gains here come from the optimizer itself.

\needspace{10\baselineskip}\subsection*{Compute accounting}
Iterations do not measure compute, so we count unit-runs: for the kernels one benchmark or gate process execution, for the optimizers one optimization of one unit (the two are not comparable). Output tokens are counted from the session transcripts, deduplicated by message identifier and including subagent transcripts.  The optimizer runs used 1.60 million unit-runs over 80 runs.

\section*{Data availability}
The data generated in this study will be made publicly available upon publication. The reaction paths are from the Birkholz--Schlegel\cite{Birkholz2015} and Baker\cite{Baker1996} sets as distributed by Marks and Gomes\cite{MarksGomes2025}. All other structures come from ASE's collections, including G2\cite{G2} and S22\cite{S22}, or from its structure builders, except three built by hand.

\section*{Code availability}
The code used in this study will be made publicly available upon publication.

\section*{Author contributions}
J.W. conceived the study, carried out all calculations and analyses, and wrote the manuscript.

\section*{Competing interests}
The author declares no competing interests.

\section*{Acknowledgements}
This work was supported by a KIAS Individual Grant (AP110201) via the Center for Artificial Intelligence and Natural Sciences at Korea Institute for Advanced Study.

\section*{Correspondence}
{\raggedright Correspondence and requests for materials should be addressed to Jeheon Woo (jhwoo1905@kias.re.kr).\par}

\bibliographystyle{naturemag}
\bibliography{refs}

\clearpage
\appendix
\section*{Supplementary Information}

\subsection*{Contents}
{\small\renewcommand{\arraystretch}{1.3}
\begin{tabularx}{\linewidth}{@{}>{\raggedright\arraybackslash}p{4.3cm}>{\raggedright\arraybackslash}X r@{}}
\hyperref[si:kernels]{Supplementary Table 1} & What the kernel runs changed & \pageref{si:kernels}\\
\hyperref[si:results]{Supplementary Table 2} & Held-out results of the optimizer runs (values of record) & \pageref{si:results}\\
\hyperref[si:termination]{Supplementary Table 3} & Forces at which converged held-out bands stop & \pageref{si:termination}\\
\hyperref[si:baselines]{Supplementary Table 4} & Post hoc checks against further stock baselines & \pageref{si:baselines}\\
\hyperref[si:overhead]{Supplementary Table 5} & Wall time inside and outside the calculator & \pageref{si:overhead}\\
\hyperref[si:autosella]{Supplementary Table 6} & Relaxers against concurrent work & \pageref{si:autosella}\\
\hyperref[si:cost]{Supplementary Table 7} & Compute and output tokens per run and configuration & \pageref{si:cost}\\
\hyperref[fig:Snopen]{Supplementary Fig. 1} & Held-out speedups without the penalty & \pageref{fig:Snopen}\\
\end{tabularx}}

\renewcommand{\thetable}{S\arabic{table}}\setcounter{table}{0}

\renewcommand{\thefigure}{\arabic{figure}}\renewcommand{\figurename}{Supplementary Fig.}\setcounter{figure}{0}

\renewcommand{\theHfigure}{S\arabic{figure}}\renewcommand{\theHtable}{S\arabic{table}}

\clearpage
\phantomsection\label{si:kernels}
\noindent Supplementary Table 1: \textbf{What the kernel runs changed.} For each kernel, where stock ASE spends its time, the change all three runs made and where the runs differ, read from the differences between stock ASE and each run's merged tree and from the agents' logs. Speedups are those of Fig. 2 (Python with NumPy and SciPy, C disabled).

\smallskip
{\scriptsize\setlength{\tabcolsep}{4pt}
\begin{tabularx}{\linewidth}{>{\hsize=0.524\hsize\raggedright\arraybackslash}X>{\hsize=1.190\hsize\raggedright\arraybackslash}X>{\hsize=1.238\hsize\raggedright\arraybackslash}X>{\hsize=1.381\hsize\raggedright\arraybackslash}X>{\hsize=0.667\hsize\raggedright\arraybackslash}X}
\toprule
\textbf{\hspace{0pt}kernel} & \textbf{\hspace{0pt}stock bottleneck} & \textbf{\hspace{0pt}main change, all three runs} & \textbf{\hspace{0pt}where the runs differ} & \textbf{\hspace{0pt}speedup, runs 1 / 2 / 3} \\ \midrule
\hspace{0pt}emt & \hspace{0pt}two per-atom Python loops: one in the calculator, one in building its neighbor list (ASE's NeighborList class) & \hspace{0pt}both replaced by array operations over one list of atom pairs, each pair stored once & \hspace{0pt}neighbor search: runs 1 and 3 use a k-d tree; run 2 computes all candidate distances in one matrix product for these small systems & \hspace{0pt}17.6 / 30.4 / 16.6 \\
\hspace{0pt}lj & \hspace{0pt}the same two loops & \hspace{0pt}a half pair list (from a k-d-tree search for the clusters and bulk cells), forces added to both atoms of each pair & \hspace{0pt}how far each leaves ASE's general neighbor list: run 1 replaces it everywhere with its own search over image copies; run 2 keeps it and reproduces its pair order; run 3 bypasses it for open cells and large orthogonal ones and computes the stress only when requested & \hspace{0pt}21.8 / 15.1 / 34.5 \\
\hspace{0pt}neighborlist & \hspace{0pt}Python loop over neighboring bins, testing every padded atom pair & \hspace{0pt}k-d-tree search over copies of the periodic images; each bond found once and mirrored & \hspace{0pt}small array-level changes; run 3 adds a periodic k-d tree for fully periodic orthogonal cells & \hspace{0pt}18.5 / 24.2 / 20.5 \\
\hspace{0pt}geometry & \hspace{0pt}the general minimum-image search (lattice reduction, then 27 translations per displacement) & \hspace{0pt}exact per-axis minimum image for orthogonal cells, used only by the two cubic cells of the nine systems & \hspace{0pt}run 2, the fastest, also tests 8 translations instead of 27 in fully periodic cells and forms all distances of an orthogonal cell at once & \hspace{0pt}12.1 / 15.4 / 12.4 \\
\hspace{0pt}analysis & \hspace{0pt}bonds and angles through a sparse connectivity matrix, all-pairs shortest paths and per-row Python loops & \hspace{0pt}bonds and angles read directly from the neighbor pairs (an angle is two bonds sharing an atom, so no shortest-path search is needed); the radial distribution in one pass over the pairs & \hspace{0pt}how duplicate bonds are removed & \hspace{0pt}21.8 / 19.5 / 19.9 \\
\hspace{0pt}idpp & \hspace{0pt}a minimum-image search per atom, which in every benchmark cell falls back to the general search & \hspace{0pt}all pairs at once, with exact per-axis rounding for orthogonal cells & \hspace{0pt}exactness: run 3 keeps the forces bit-identical to stock ASE (true powers and divisions, and stock's exact minimum image) and is 2.7-fold slower than run 1 & \hspace{0pt}99.4 / 58.5 / 37.4 \\
\bottomrule
\end{tabularx}
\normalsize}
\smallskip

\clearpage
\phantomsection\label{si:results}
\noindent Supplementary Table 2: \textbf{Held-out results of the optimizer runs (values of record).} Mean force calls per held-out unit of each task's finals and parameter floor, with 500 (relaxation) or 7,009 (paths) calls charged per unconverged unit; bold, the speedup of the mean of runs over the reference.

\smallskip
{\tiny\setlength{\tabcolsep}{4pt}
\begin{tabularx}{\linewidth}{>{\hsize=1.038\hsize\raggedright\arraybackslash}X>{\hsize=1.038\hsize\raggedright\arraybackslash}X>{\hsize=0.890\hsize\raggedright\arraybackslash}X>{\hsize=0.692\hsize\raggedright\arraybackslash}X>{\hsize=0.989\hsize\raggedright\arraybackslash}X>{\hsize=1.137\hsize\raggedright\arraybackslash}X>{\hsize=1.038\hsize\raggedright\arraybackslash}X>{\hsize=1.236\hsize\raggedright\arraybackslash}X>{\hsize=0.940\hsize\raggedright\arraybackslash}X}
\toprule
\textbf{\hspace{0pt}task} & \textbf{\hspace{0pt}protocol (model)} & \textbf{\hspace{0pt}reference (stock ASE)} & \textbf{\hspace{0pt}reference mean (calls per unit)} & \textbf{\hspace{0pt}parameter floor: mean, speedup (units converged)} & \textbf{\hspace{0pt}finals: calls per unit, mean (range over runs)} & \textbf{\hspace{0pt}speedup (range over runs)} & \textbf{\hspace{0pt}units converged} & \textbf{\hspace{0pt}speedup on units both converge, no penalty (range)} \\ \midrule
\multicolumn{9}{l}{\textbf{Structure relaxation (30 held-out units)}} \\ \addlinespace[1pt]
\hspace{0pt}BFGS & \hspace{0pt}single-phase + two-phase (Opus 5) & \hspace{0pt}BFGS & \hspace{0pt}50.6 & \hspace{0pt}34.4, 1.47$\times$ (30/30) & \hspace{0pt}26.5 (17.9--33.8) & \hspace{0pt}\textbf{1.91$\times$} (1.49--2.82$\times$) & \hspace{0pt}all 30 & \hspace{0pt}same \\
\hspace{0pt}FIRE & \hspace{0pt}single-phase + two-phase (Opus 5) & \hspace{0pt}FIRE & \hspace{0pt}94.3 & \hspace{0pt}45.0, 2.09$\times$ (30/30) & \hspace{0pt}48.0 (43.1--54.9) & \hspace{0pt}\textbf{1.96$\times$} (1.72--2.19$\times$) & \hspace{0pt}all 30 & \hspace{0pt}same \\
\hspace{0pt}LBFGS & \hspace{0pt}single-phase + two-phase (Opus 5) & \hspace{0pt}LBFGS & \hspace{0pt}50.5 & \hspace{0pt}34.4, 1.47$\times$ (30/30) & \hspace{0pt}29.9 (24.8--31.4) & \hspace{0pt}\textbf{1.69$\times$} (1.61--2.03$\times$) & \hspace{0pt}all 30 & \hspace{0pt}same \\
\hspace{0pt}BFGS\-Line\-Search & \hspace{0pt}single-phase + two-phase (Opus 5) & \hspace{0pt}BFGS\-Line\-Search & \hspace{0pt}38.5 & \hspace{0pt}36.4, 1.06$\times$ (30/30) & \hspace{0pt}35.3 (30.2--46.5) & \hspace{0pt}\textbf{1.09$\times$} (0.83--1.28$\times$) & \hspace{0pt}30/30, 30/30, 30/30, 30/30, 30/30, 29/30 & \hspace{0pt}1.10--1.28$\times$ \\
\hspace{0pt}relaxer from scratch & \hspace{0pt}single-phase + two-phase (Opus 5) & \hspace{0pt}LBFGS & \hspace{0pt}50.5 & \hspace{0pt}268.3, 0.19$\times$ (21/30) & \hspace{0pt}30.6 (24.5--45.1) & \hspace{0pt}\textbf{1.65$\times$} (1.12--2.06$\times$) & \hspace{0pt}30/30, 30/30, 30/30, 29/30, 30/30, 30/30 & \hspace{0pt}1.62--2.06$\times$ \\
\hspace{0pt}relaxer from scratch & \hspace{0pt}single-phase (Sonnet 5) & \hspace{0pt}LBFGS & \hspace{0pt}50.5 & \hspace{0pt}--- & \hspace{0pt}38.1 (35.7--39.3) & \hspace{0pt}\textbf{1.33$\times$} (1.28--1.42$\times$) & \hspace{0pt}all 30 & \hspace{0pt}same \\
\hspace{0pt}relaxer from scratch & \hspace{0pt}single-phase (Fable 5.1) & \hspace{0pt}LBFGS & \hspace{0pt}50.5 & \hspace{0pt}--- & \hspace{0pt}29.1 (22.0--33.3) & \hspace{0pt}\textbf{1.74$\times$} (1.52--2.30$\times$) & \hspace{0pt}all 30 & \hspace{0pt}same \\
\hspace{0pt}relaxer from scratch & \hspace{0pt}informed (Opus 5) & \hspace{0pt}LBFGS & \hspace{0pt}50.5 & \hspace{0pt}--- & \hspace{0pt}17.1 (16.9--17.2) & \hspace{0pt}\textbf{2.96$\times$} (2.94--2.99$\times$) & \hspace{0pt}all 30 & \hspace{0pt}same \\
\hspace{0pt}relaxer from scratch & \hspace{0pt}informed (Fable 5.1) & \hspace{0pt}LBFGS & \hspace{0pt}50.5 & \hspace{0pt}--- & \hspace{0pt}15.7 (15.0--16.4) & \hspace{0pt}\textbf{3.22$\times$} (3.07--3.37$\times$) & \hspace{0pt}all 30 & \hspace{0pt}same \\
\addlinespace[3pt]\multicolumn{9}{l}{\textbf{Reaction paths (19 held-out reactions)}} \\ \addlinespace[1pt]
\hspace{0pt}CI-NEB+\allowbreak{}FIRE & \hspace{0pt}single-phase + two-phase (Opus 5) & \hspace{0pt}CI-NEB+\allowbreak{}FIRE & \hspace{0pt}2,109.7 & \hspace{0pt}2,292.5, 0.92$\times$ (15/19) & \hspace{0pt}1,040.0 (641.3--1,631.1) & \hspace{0pt}\textbf{2.03$\times$} (1.29--3.29$\times$) & \hspace{0pt}19/19, 18/19, 16/19, 18/19, 18/19, 19/19 & \hspace{0pt}1.5--3.1$\times$ \\
\hspace{0pt}CI-NEB+\allowbreak{}ODE12r & \hspace{0pt}single-phase + two-phase (Opus 5) & \hspace{0pt}CI-NEB+\allowbreak{}ODE12r & \hspace{0pt}2,236.1 & \hspace{0pt}937.8, 2.38$\times$ (18/19) & \hspace{0pt}903.3 (517.1--1,371.6) & \hspace{0pt}\textbf{2.48$\times$} (1.63--4.32$\times$) & \hspace{0pt}19/19, 18/19, 17/19, 16/19, 18/19, 19/19 & \hspace{0pt}1.8--4.5$\times$ \\
\hspace{0pt}path optimizer from scratch & \hspace{0pt}single-phase + two-phase (Opus 5) & \hspace{0pt}CI-NEB+\allowbreak{}ODE12r & \hspace{0pt}2,236.1 & \hspace{0pt}4,278.6, 0.52$\times$ (11/19) & \hspace{0pt}1,128.8 (772.2--1,514.7) & \hspace{0pt}\textbf{1.98$\times$} (1.48--2.90$\times$) & \hspace{0pt}17/19, 16/19, 18/19, 17/19, 17/19, 17/19 & \hspace{0pt}1.6--3.6$\times$ \\
\hspace{0pt}path optimizer from scratch & \hspace{0pt}single-phase (Sonnet 5) & \hspace{0pt}CI-NEB+\allowbreak{}ODE12r & \hspace{0pt}2,236.1 & \hspace{0pt}--- & \hspace{0pt}2,019.0 (1,595.8--2,376.5) & \hspace{0pt}\textbf{1.11$\times$} (0.94--1.40$\times$) & \hspace{0pt}14/19, 18/19, 18/19 & \hspace{0pt}0.7--1.4$\times$ \\
\hspace{0pt}path optimizer from scratch & \hspace{0pt}single-phase (Fable 5.1) & \hspace{0pt}CI-NEB+\allowbreak{}ODE12r & \hspace{0pt}2,236.1 & \hspace{0pt}--- & \hspace{0pt}602.1 (400.1--887.6) & \hspace{0pt}\textbf{3.71$\times$} (2.52--5.59$\times$) & \hspace{0pt}19/19, 18/19, 19/19 & \hspace{0pt}1.8--4.3$\times$ \\
\hspace{0pt}path optimizer from scratch & \hspace{0pt}informed (Opus 5) & \hspace{0pt}CI-NEB+\allowbreak{}ODE12r & \hspace{0pt}2,236.1 & \hspace{0pt}--- & \hspace{0pt}450.2 (156.5--935.8) & \hspace{0pt}\textbf{4.97$\times$} (2.39--14.29$\times$) & \hspace{0pt}19/19, 17/19, 19/19 & \hspace{0pt}5.2--9.3$\times$ \\
\hspace{0pt}path optimizer from scratch & \hspace{0pt}informed (Fable 5.1) & \hspace{0pt}CI-NEB+\allowbreak{}ODE12r & \hspace{0pt}2,236.1 & \hspace{0pt}--- & \hspace{0pt}256.4 (127.9--481.1) & \hspace{0pt}\textbf{8.72$\times$} (4.65--17.48$\times$) & \hspace{0pt}19/19, 19/19, 18/19 & \hspace{0pt}7.5--12.0$\times$ \\
\bottomrule
\end{tabularx}
\normalsize}
\smallskip

{\footnotesize\begin{itemize}\setlength{\itemsep}{1pt}
\item Speedup = reference mean / run mean, both in force calls per held-out unit; ``mean of runs'' divides the reference mean by the mean of the run means, and the ranges give the individual runs.
\item References: library start, the stock method being optimized; from scratch and informed, L-BFGS (relaxation) and CI-NEB+ODE12r (paths), the stock methods with the fewest force calls on the optimization set. Both stock path references converge 16 of 19 reactions; every stock relaxation reference converges all 30 units.
\item ``Units both converge'' averages, without penalty, over the units that both the run and its reference converge. In relaxation, two finals each leave one unit unconverged (two-phase run 3 on BFGSLineSearch, two-phase run 1 from scratch). Every other final converges all 30, so the two views coincide.
\item Parameter floor: each task's parameter-only run (Methods), evaluated on the held-out units like every final; for the tasks from scratch, the tuned starting optimizer. The starting optimizers themselves need 239.0 force calls per relaxation unit (23/30 converged) and 3,498.3 per reaction (15/19); the floors from scratch and on CI-NEB+FIRE need more force calls than their starting points.
\item Cross-task aggregate of the single-phase and two-phase finals (Opus 5): the geometric mean over the eight tasks of the mean-of-runs speedup is 1.81$\times$ (relaxation 1.63$\times$, paths 2.15$\times$).
\item Stock LBFGS (memory 100) and stock BFGS differ by one call on two held-out units, and their parameter floors settled on the same values (initial curvature 20, maximum step 0.12 \AA{}), with identical per-unit counts.
\item ``Units converged'', where listed by run, gives the runs in the order single-phase runs 1, 2, 3, then two-phase runs 1, 2, 3; the informed, Sonnet 5 and Fable 5.1 rows list runs 1, 2, 3.
\end{itemize}}

\clearpage
\phantomsection\label{si:termination}
\noindent Supplementary Table 3: \textbf{Forces at which converged held-out bands stop.} The two forces the held-out gate checks, re-evaluated with fresh calculators on every band an optimizer declared converged on the 19 held-out reactions. Every optimizer was called with the same 0.05 eV/\AA{} convergence threshold; the gate accepts at most 0.15 eV/\AA{} and applies the in-loop path checks (Methods). The stock references were re-run to 3,000 steps, and bands they converged only after the metric's 1,000-step cap are excluded because the metric counts them as unconverged.

\smallskip
{\scriptsize\setlength{\tabcolsep}{4pt}
\begin{tabularx}{\linewidth}{>{\hsize=1.969\hsize\raggedright\arraybackslash}X>{\hsize=0.375\hsize\raggedright\arraybackslash}X>{\hsize=0.562\hsize\raggedright\arraybackslash}X>{\hsize=1.266\hsize\raggedright\arraybackslash}X>{\hsize=1.266\hsize\raggedright\arraybackslash}X>{\hsize=0.562\hsize\raggedright\arraybackslash}X}
\toprule
\textbf{\hspace{0pt}optimizers} & \textbf{\hspace{0pt}runs} & \textbf{\hspace{0pt}bands declared converged} & \textbf{\hspace{0pt}true perpendicular force, largest interior image of each band: median / 90th percentile / max over bands (eV/\AA{})} & \textbf{\hspace{0pt}full force on the highest-energy image: median / 90th percentile / max over bands (eV/\AA{})} & \textbf{\hspace{0pt}refused by the gate} \\ \midrule
\hspace{0pt}stock CI-NEB+\allowbreak{}FIRE & \hspace{0pt}1 & \hspace{0pt}16 & \hspace{0pt}0.043 / 0.050 / 0.050 & \hspace{0pt}0.037 / 0.050 / 0.053 & \hspace{0pt}0 \\
\hspace{0pt}stock CI-NEB+\allowbreak{}ODE12r & \hspace{0pt}1 & \hspace{0pt}16 & \hspace{0pt}0.052 / 0.064 / 0.067 & \hspace{0pt}0.016 / 0.055 / 0.067 & \hspace{0pt}0 \\
\hspace{0pt}library finals (CI-NEB+\allowbreak{}FIRE, CI-NEB+\allowbreak{}ODE12r) & \hspace{0pt}12 & \hspace{0pt}216 & \hspace{0pt}0.050 / 0.067 / 0.144 & \hspace{0pt}0.044 / 0.066 / 0.163 & \hspace{0pt}1 \\
\hspace{0pt}from-scratch finals (all models) & \hspace{0pt}12 & \hspace{0pt}212 & \hspace{0pt}0.047 / 0.050 / 0.054 & \hspace{0pt}0.041 / 0.051 / 0.193 & \hspace{0pt}4 \\
\hspace{0pt}informed finals (Fable 5.1, Opus 5) & \hspace{0pt}6 & \hspace{0pt}112 & \hspace{0pt}0.048 / 0.050 / 0.054 & \hspace{0pt}0.041 / 0.049 / 0.060 & \hspace{0pt}1 \\
\bottomrule
\end{tabularx}
\normalsize}
\smallskip

\clearpage
\phantomsection\label{si:baselines}
\noindent Supplementary Table 4: \textbf{Post hoc checks against further stock baselines.} Post hoc checks of the references (Methods) against further stock options: A, Sella (relaxation); B, ASE's preconditioned band optimizer (paths); C, ASE's preconditioned relaxer (relaxation).

\needspace{8\baselineskip}\medskip\noindent \textbf{A.\ Relaxation: Sella.} Sella 2.5.0\cite{Sella}, set to minimize (\texttt{order=0}), in two configurations: package defaults (Cartesian coordinates) and internal coordinates. With internal coordinates, Sella did not finish the nine metal units within the one-hour job limit; those units are scored with the default configuration, and the baseline takes Sella's better configuration on every unit.

\needspace{8\baselineskip}\medskip\noindent \textbf{Sella} (mean force calls):

\smallskip
{\scriptsize
\begin{tabularx}{\linewidth}{>{\hsize=1.600\hsize\raggedright\arraybackslash}X>{\hsize=0.700\hsize\raggedright\arraybackslash}X>{\hsize=0.800\hsize\raggedright\arraybackslash}X>{\hsize=0.900\hsize\raggedright\arraybackslash}X}
\toprule
\textbf{\hspace{0pt}} & \textbf{\hspace{0pt}all 30} & \textbf{\hspace{0pt}molecules (21)} & \textbf{\hspace{0pt}metals (9: Pd, Ag bulk; Au$_{55}$)} \\ \midrule
\hspace{0pt}default & \hspace{0pt}33.73 & \hspace{0pt}35.33 & \hspace{0pt}30.00 \\
\hspace{0pt}internal coordinates & \hspace{0pt}--- & \hspace{0pt}27.19 & \hspace{0pt}did not finish \\
\hspace{0pt}better of the two, per unit & \hspace{0pt}27.47 & \hspace{0pt}26.38 & \hspace{0pt}30.00 \\
\hspace{0pt}stock LBFGS, for comparison & \hspace{0pt}50.50 & \hspace{0pt}47.43 & \hspace{0pt}57.67 \\
\bottomrule
\end{tabularx}
\normalsize}
\smallskip

\needspace{8\baselineskip}\medskip\noindent \textbf{Relaxer finals against the better of the two} (ratio = Sella mean / final mean):

\smallskip
{\scriptsize\setlength{\tabcolsep}{4pt}
\begin{tabularx}{\linewidth}{>{\hsize=2.094\hsize\raggedright\arraybackslash}X>{\hsize=0.777\hsize\raggedright\arraybackslash}X>{\hsize=0.777\hsize\raggedright\arraybackslash}X>{\hsize=0.777\hsize\raggedright\arraybackslash}X>{\hsize=0.576\hsize\raggedright\arraybackslash}X}
\toprule
\textbf{\hspace{0pt}finals} & \textbf{\hspace{0pt}all 30} & \textbf{\hspace{0pt}molecules} & \textbf{\hspace{0pt}metals} & \textbf{\hspace{0pt}units better than Sella (of 30)} \\ \midrule
\hspace{0pt}informed, Fable 5.1 (runs 1 / 2 / 3) & \hspace{0pt}1.76 / 1.83 / 1.67 & \hspace{0pt}1.76 / 1.87 / 1.56 & \hspace{0pt}1.74 / 1.75 / 1.94 & \hspace{0pt}27 / 27 / 30 \\
\hspace{0pt}informed, Opus 5 (runs 1 / 2 / 3) & \hspace{0pt}1.60 / 1.60 / 1.63 & \hspace{0pt}1.54 / 1.59 / 1.68 & \hspace{0pt}1.72 / 1.64 / 1.53 & \hspace{0pt}26 / 26 / 29 \\
\hspace{0pt}single-phase from scratch, Opus 5 (runs 1--3) & \hspace{0pt}0.95--1.12 & \hspace{0pt}0.91--1.00 & \hspace{0pt}1.03--1.51 & \hspace{0pt}10--18 \\
\hspace{0pt}single-phase from scratch, Fable 5.1 (runs 1--3) & \hspace{0pt}0.83--1.25 & \hspace{0pt}0.68--1.10 & \hspace{0pt}0.70--1.72 & \hspace{0pt}9--17 \\
\hspace{0pt}single-phase from scratch, Sonnet 5 (runs 1--3) & \hspace{0pt}0.70--0.77 & \hspace{0pt}0.61--0.75 & \hspace{0pt}0.61--1.19 & \hspace{0pt}6--7 \\
\bottomrule
\end{tabularx}
\normalsize}
\smallskip

{\footnotesize On the two aromatic dimers (phenol dimer, pyrazine dimer), the Cartesian optimizers, ASE's BFGS and L-BFGS included, stop up to 14 meV above Sella's internal-coordinate run at the same force threshold; on the other molecules all agree within 2 meV. The force-only criterion counts the dimer units as converged for every optimizer, but there the call counts are not compared at equal final energy.\par}

\needspace{8\baselineskip}\medskip\noindent \textbf{B.\ Reaction paths: preconditioned band optimizer.} The preconditioned NEB of Makri et al.\cite{Makri2019} as shipped in ASE (per-image \texttt{Exp} preconditioner with adaptive ODE12r integration, \texttt{NEBOptimizer(method=\textquotesingle{}ODE\textquotesingle{})}), was run with the spline and string tangents (the two for which ASE provides the preconditioner) and with the spline tangent unpreconditioned; the baseline takes the best of the three on each reaction. All runs use ASE defaults (spring constant 0.1 eV/\AA{}$^{2}$, \texttt{Exp} parameters estimated automatically) except the climbing image, which is on from the start, as in every CI-NEB evaluation here.

\smallskip
{\scriptsize\setlength{\tabcolsep}{4pt}
\begin{tabularx}{\linewidth}{>{\hsize=1.958\hsize\raggedright\arraybackslash}X>{\hsize=0.958\hsize\raggedright\arraybackslash}X>{\hsize=0.667\hsize\raggedright\arraybackslash}X>{\hsize=0.792\hsize\raggedright\arraybackslash}X>{\hsize=0.625\hsize\raggedright\arraybackslash}X}
\toprule
\textbf{\hspace{0pt}method} & \textbf{\hspace{0pt}mean force calls, pre-specified (gated)} & \textbf{\hspace{0pt}solved} & \textbf{\hspace{0pt}mean force calls, own stopping test only} & \textbf{\hspace{0pt}converged by own test} \\ \midrule
\hspace{0pt}CI-NEB+\allowbreak{}FIRE (default) & \hspace{0pt}2,109.7 & \hspace{0pt}16 / 19 & \hspace{0pt}2,109.7 & \hspace{0pt}16 / 19 \\
\hspace{0pt}CI-NEB+\allowbreak{}ODE12r & \hspace{0pt}2,236.1 & \hspace{0pt}16 / 19 & \hspace{0pt}2,236.1 & \hspace{0pt}16 / 19 \\
\hspace{0pt}spline + \texttt{Exp} + ODE12r & \hspace{0pt}6,344.9 & \hspace{0pt}2 / 19 & \hspace{0pt}6,034.8 & \hspace{0pt}3 / 19 \\
\hspace{0pt}string + \texttt{Exp} + ODE12r & \hspace{0pt}6,365.6 & \hspace{0pt}2 / 19 & \hspace{0pt}4,639.2 & \hspace{0pt}9 / 19 \\
\hspace{0pt}spline + ODE12r (no preconditioner) & \hspace{0pt}6,058.5 & \hspace{0pt}3 / 19 & \hspace{0pt}3,761.7 & \hspace{0pt}10 / 19 \\
\hspace{0pt}best of the three above, per reaction & \hspace{0pt}5,753.6 & \hspace{0pt}4 / 19 & \hspace{0pt}2,861.0 & \hspace{0pt}13 / 19 \\
\hspace{0pt}informed finals, Fable 5.1 (runs 1 / 2 / 3) & \hspace{0pt}127.9 / 160.2 / 481.1 & \hspace{0pt}19, 19, 18 / 19 & \hspace{0pt} & \hspace{0pt} \\
\bottomrule
\end{tabularx}
\normalsize}
\smallskip

{\footnotesize At its defaults the preconditioned band optimizer was not a stronger reference on these reactions, with or without the gate. Most failures (29 of 50) are runs that ODE12r stopped without converging, after 319--5,329 force calls, well short of the 1,000-step cap. Bands that reached the cap, which ASE's ODE12r reports as converged, count as failures in both columns (two spline and three string bands); one spline band refused by the path-resolution check counts as a failure in the gated columns. For the spline and string drivers, the gate (which measures the perpendicular force with the improved tangent in the Euclidean metric, as for every agent) is stricter than the driver's own test; the right-hand columns, without the gate, bound that effect.\par}

\needspace{8\baselineskip}\medskip\noindent \textbf{C.\ Relaxation: ASE's preconditioned relaxer.} ASE's \texttt{PreconLBFGS}\cite{Packwood2016}, at its defaults and with the \texttt{Exp} preconditioner forced on.

\smallskip
{\scriptsize
\begin{tabularx}{\linewidth}{>{\hsize=2.122\hsize\raggedright\arraybackslash}X>{\hsize=0.735\hsize\raggedright\arraybackslash}X>{\hsize=0.449\hsize\raggedright\arraybackslash}X>{\hsize=0.694\hsize\raggedright\arraybackslash}X}
\toprule
\textbf{\hspace{0pt}configuration} & \textbf{\hspace{0pt}mean force calls (500 per failure)} & \textbf{\hspace{0pt}converged} & \textbf{\hspace{0pt}mean, converged units only} \\ \midrule
\hspace{0pt}LBFGS (default, reference) & \hspace{0pt}50.5 & \hspace{0pt}30 / 30 & \hspace{0pt}50.5 \\
\hspace{0pt}\texttt{PreconLBFGS}, defaults (preconditioner off below 100 atoms) & \hspace{0pt}119.5 & \hspace{0pt}29 / 30 & \hspace{0pt}106.4 \\
\hspace{0pt}\texttt{PreconLBFGS}, \texttt{Exp} forced on & \hspace{0pt}421.9 & \hspace{0pt}5 / 30 & \hspace{0pt}31.4 \\
\hspace{0pt}\texttt{PreconLBFGS}, \texttt{Exp}, non-periodic units placed in a box & \hspace{0pt}92.5 & \hspace{0pt}29 / 30 & \hspace{0pt}78.4 \\
\bottomrule
\end{tabularx}
\normalsize}
\smallskip

{\footnotesize With \texttt{Exp} forced on, the preconditioner cannot be set up on the 24 non-periodic units as built (molecules, dimers and Au$_{55}$; ASE's nearest-neighbor estimate fails), so these were also run in a box (a rigid translation plus a cell that the non-periodic calculator ignores). On the 29 units it converges, boxed \texttt{Exp} needs 78.4 calls, against 15.1--16.6 for the informed Fable 5.1 finals and 17.0--17.3 for Opus 5. At its defaults or with \texttt{Exp} on, ASE's preconditioned relaxer is not a stronger reference than L-BFGS on these units.\par}

\clearpage
\phantomsection\label{si:overhead}
\noindent Supplementary Table 5: \textbf{Wall time inside and outside the calculator.} Wall time of the optimization per held-out relaxation unit (means over the units; the setup of structure and calculator excluded), split into the time inside the calculator, which the metric counts as force evaluations, and the rest, the optimizer's own work, which it does not count. Each optimizer ran through the held-out worker of record with timers around the counted calculator call and the setup, at 0.20 \AA{} displacements; all units ran on one node, 16 at a time, one core per unit. One GFN1-xTB evaluation takes 29 ms on the molecules and 4.0 s on the metal units (Pd and Ag crystals, Au$_{55}$ cluster) on average.

\smallskip
{\tiny\setlength{\tabcolsep}{4pt}
\begin{tabularx}{\linewidth}{>{\hsize=2.133\hsize\raggedright\arraybackslash}X>{\hsize=0.988\hsize\raggedright\arraybackslash}X>{\hsize=0.728\hsize\raggedright\arraybackslash}X>{\hsize=0.832\hsize\raggedright\arraybackslash}X>{\hsize=0.936\hsize\raggedright\arraybackslash}X>{\hsize=0.728\hsize\raggedright\arraybackslash}X>{\hsize=0.832\hsize\raggedright\arraybackslash}X>{\hsize=0.832\hsize\raggedright\arraybackslash}X>{\hsize=0.988\hsize\raggedright\arraybackslash}X}
\toprule
\textbf{\hspace{0pt}optimizer} & \textbf{\hspace{0pt}molecules (21): force calls} & \textbf{\hspace{0pt}wall time (s)} & \textbf{\hspace{0pt}outside the calculator} & \textbf{\hspace{0pt}metal units (9): force calls} & \textbf{\hspace{0pt}wall time (s)} & \textbf{\hspace{0pt}outside the calculator} & \textbf{\hspace{0pt}all 30: wall time (s)} & \textbf{\hspace{0pt}LBFGS / this optimizer, wall time} \\ \midrule
\hspace{0pt}LBFGS (stock) & \hspace{0pt}47.4 & \hspace{0pt}1.63 & \hspace{0pt}8\,\% & \hspace{0pt}57.7 & \hspace{0pt}210.5 & \hspace{0pt}0.1\,\% & \hspace{0pt}64.3 & \hspace{0pt}1.00 \\
\hspace{0pt}BFGS (stock) & \hspace{0pt}47.5 & \hspace{0pt}1.65 & \hspace{0pt}9\,\% & \hspace{0pt}57.8 & \hspace{0pt}211.4 & \hspace{0pt}0.2\,\% & \hspace{0pt}64.6 & \hspace{0pt}1.00 \\
\hspace{0pt}BFGS\-Line\-Search (stock) & \hspace{0pt}44.2 & \hspace{0pt}1.67 & \hspace{0pt}8\,\% & \hspace{0pt}25.2 & \hspace{0pt}99.9 & \hspace{0pt}0.1\,\% & \hspace{0pt}31.2 & \hspace{0pt}2.06 \\
\hspace{0pt}FIRE (stock) & \hspace{0pt}99.7 & \hspace{0pt}2.73 & \hspace{0pt}8\,\% & \hspace{0pt}81.6 & \hspace{0pt}340.8 & \hspace{0pt}0.1\,\% & \hspace{0pt}104.2 & \hspace{0pt}0.62 \\
\hspace{0pt}informed, Fable 5.1, run 1 & \hspace{0pt}15.0 & \hspace{0pt}0.42 & \hspace{0pt}15\,\% & \hspace{0pt}17.2 & \hspace{0pt}68.4 & \hspace{0pt}0.5\,\% & \hspace{0pt}20.8 & \hspace{0pt}3.09 \\
\hspace{0pt}informed, Fable 5.1, run 2 & \hspace{0pt}14.1 & \hspace{0pt}0.53 & \hspace{0pt}40\,\% & \hspace{0pt}17.1 & \hspace{0pt}93.5 & \hspace{0pt}23.6\,\% & \hspace{0pt}28.4 & \hspace{0pt}2.26 \\
\hspace{0pt}informed, Fable 5.1, run 3 & \hspace{0pt}16.9 & \hspace{0pt}0.74 & \hspace{0pt}33\,\% & \hspace{0pt}15.4 & \hspace{0pt}68.0 & \hspace{0pt}3.9\,\% & \hspace{0pt}20.9 & \hspace{0pt}3.07 \\
\hspace{0pt}informed, Opus 5, run 1 & \hspace{0pt}17.1 & \hspace{0pt}0.86 & \hspace{0pt}43\,\% & \hspace{0pt}17.4 & \hspace{0pt}78.6 & \hspace{0pt}6.2\,\% & \hspace{0pt}24.2 & \hspace{0pt}2.66 \\
\hspace{0pt}informed, Opus 5, run 2 & \hspace{0pt}16.6 & \hspace{0pt}0.70 & \hspace{0pt}34\,\% & \hspace{0pt}18.3 & \hspace{0pt}81.7 & \hspace{0pt}5.8\,\% & \hspace{0pt}25.0 & \hspace{0pt}2.57 \\
\hspace{0pt}informed, Opus 5, run 3 & \hspace{0pt}15.7 & \hspace{0pt}0.52 & \hspace{0pt}16\,\% & \hspace{0pt}19.7 & \hspace{0pt}81.3 & \hspace{0pt}0.6\,\% & \hspace{0pt}24.7 & \hspace{0pt}2.60 \\
\bottomrule
\end{tabularx}
\normalsize}
\smallskip

\clearpage
\phantomsection\label{si:autosella}
\noindent Supplementary Table 6: \textbf{Relaxers against concurrent work.} Post hoc comparison of the relaxers with the three optimizers published with concurrent work\cite{AutoSella}, each the final of one autoresearch run that started from Sella 2.5.0: AutoSella-F (Claude Fable 5.1 under Claude Code), AutoSella-A (GPT-6 Astra under Codex) and AutoSella-G (Grok 4.6). Part A uses our held-out molecules; part B uses that work's benchmark and harness.

\needspace{8\baselineskip}\medskip\noindent \textbf{A.\ Our held-out molecules.} The 21 molecular held-out relaxation units (7 systems $\times$ 3 displacement seeds). The AutoSella optimizers were called through their own entry point (code at commit b964218e, with its pinned JAX version), which takes no cell, so the periodic metal units were not run; Au$_{55}$, which Sella in internal coordinates did not finish (Supplementary Table 4, part A), was left out with them. That code's Sella reference, run the same way, gave 27.10 calls against 27.19 for our Sella run.

\needspace{8\baselineskip}\medskip\noindent \textbf{Mean force calls on the 21 molecular units} (every optimizer converged all 21):

\smallskip
{\scriptsize
\begin{tabularx}{\linewidth}{>{\hsize=1.800\hsize\raggedright\arraybackslash}X>{\hsize=0.600\hsize\raggedright\arraybackslash}X>{\hsize=0.600\hsize\raggedright\arraybackslash}X}
\toprule
\textbf{\hspace{0pt}optimizer} & \textbf{\hspace{0pt}mean force calls} & \textbf{\hspace{0pt}AutoSella-A / this optimizer} \\ \midrule
\hspace{0pt}LBFGS (stock ASE) & \hspace{0pt}47.43 & \hspace{0pt}0.37 \\
\hspace{0pt}Sella 2.5.0, internal coordinates & \hspace{0pt}27.19 & \hspace{0pt}0.64 \\
\hspace{0pt}AutoSella-G & \hspace{0pt}27.10 & \hspace{0pt}0.64 \\
\hspace{0pt}AutoSella-F & \hspace{0pt}28.57 & \hspace{0pt}0.61 \\
\hspace{0pt}AutoSella-A & \hspace{0pt}17.33 & \hspace{0pt}1 \\
\hspace{0pt}informed, Fable 5.1 (runs 1 / 2 / 3) & \hspace{0pt}14.95 / 14.10 / 16.86 & \hspace{0pt}1.16 / 1.23 / 1.03 \\
\hspace{0pt}informed, Opus 5 (runs 1 / 2 / 3) & \hspace{0pt}17.10 / 16.62 / 15.71 & \hspace{0pt}1.01 / 1.04 / 1.10 \\
\hspace{0pt}single-phase from scratch (nine runs: Opus 5, Fable 5.1, Sonnet 5) & \hspace{0pt}23.90--43.05 & \hspace{0pt}0.40--0.73 \\
\bottomrule
\end{tabularx}
\normalsize}
\smallskip

{\footnotesize The best AutoSella variant on each unit averages 16.76 calls. All three AutoSella optimizers end within 0.5 meV of Sella on every unit. On the two aromatic dimers (six units), the Cartesian optimizers, the relaxers included, stop up to 14 meV higher (Supplementary Table 4, part A), so there the counts are not compared at equal final energy.\par}

\needspace{8\baselineskip}\medskip\noindent \textbf{B.\ The benchmark of concurrent work.} That work's harness and analyzer, unchanged (branch \texttt{xtb\_ff\_evaluation}, commit 5619cd35): GFN2-xTB, a five-criterion convergence test, 200 force calls per molecule (500 for the solvated set), on four sets of drug-like and biomolecular structures from conformer generators, started far from their minima. Seven relaxer finals ran unchanged and untuned through a wrapper that lets the harness's convergence test stop them and passes its force threshold to their near-convergence damping. Sella 2.5.0, AutoSella-F and AutoSella-A ran the same way and reproduced AutoSella's published ratios within 1.2 percentage points.

\needspace{8\baselineskip}\medskip\noindent \textbf{Force calls relative to Sella} (pooled over molecules where both converged; lower is better), with the number of molecules not converged in parentheses (Sella: none):

\smallskip
{\scriptsize\setlength{\tabcolsep}{4pt}
\begin{tabularx}{\linewidth}{>{\hsize=2.043\hsize\raggedright\arraybackslash}X>{\hsize=0.652\hsize\raggedright\arraybackslash}X>{\hsize=0.739\hsize\raggedright\arraybackslash}X>{\hsize=0.870\hsize\raggedright\arraybackslash}X>{\hsize=0.696\hsize\raggedright\arraybackslash}X}
\toprule
\textbf{\hspace{0pt}optimizer} & \textbf{\hspace{0pt}test (500)} & \textbf{\hspace{0pt}dipeptides (673)} & \textbf{\hspace{0pt}amino-acid--ligand pairs (475)} & \textbf{\hspace{0pt}solvated (100)} \\ \midrule
\hspace{0pt}AutoSella-F & \hspace{0pt}72\,\% (0) & \hspace{0pt}64\,\% (0) & \hspace{0pt}68\,\% (0) & \hspace{0pt}38\,\% (0) \\
\hspace{0pt}AutoSella-A & \hspace{0pt}80\,\% (0) & \hspace{0pt}77\,\% (0) & \hspace{0pt}91\,\% (1) & \hspace{0pt}112\,\% (0) \\
\hspace{0pt}informed, Fable 5.1, run 1 & \hspace{0pt}257\,\% (6) & \hspace{0pt}255\,\% (31) & \hspace{0pt}221\,\% (175) & \hspace{0pt}132\,\% (8) \\
\hspace{0pt}informed, Fable 5.1, run 2 & \hspace{0pt}404\,\% (126) & \hspace{0pt}375\,\% (300) & \hspace{0pt}268\,\% (452) & \hspace{0pt}450\,\% (99) \\
\hspace{0pt}informed, Fable 5.1, run 3 & \hspace{0pt}218\,\% (5) & \hspace{0pt}201\,\% (6) & \hspace{0pt}216\,\% (91) & \hspace{0pt}119\,\% (3) \\
\hspace{0pt}informed, Opus 5, run 1 & \hspace{0pt}230\,\% (5) & \hspace{0pt}216\,\% (10) & \hspace{0pt}224\,\% (146) & \hspace{0pt}120\,\% (2) \\
\hspace{0pt}informed, Opus 5, run 2 & \hspace{0pt}270\,\% (13) & \hspace{0pt}251\,\% (33) & \hspace{0pt}217\,\% (199) & \hspace{0pt}147\,\% (16) \\
\hspace{0pt}informed, Opus 5, run 3 & \hspace{0pt}292\,\% (16) & \hspace{0pt}291\,\% (45) & \hspace{0pt}233\,\% (164) & \hspace{0pt}128\,\% (9) \\
\hspace{0pt}single-phase from scratch, Opus 5, run 1 & \hspace{0pt}309\,\% (10) & \hspace{0pt}288\,\% (29) & \hspace{0pt}289\,\% (247) & \hspace{0pt}143\,\% (5) \\
\bottomrule
\end{tabularx}
\normalsize}
\smallskip

{\footnotesize The relaxers reach Sella's minimum (heavy-atom RMSD below 0.1 \AA{} and energy within 0.1 kcal/mol) on 37--84\,\% of the test molecules, 13--83\,\% of the dipeptides and at most 19\,\% of the other two sets, and otherwise stop at higher energy; the AutoSella optimizers reach it on 93--97\,\% of the test and dipeptide molecules.\par}

{\footnotesize Our optimization and held-out sets hold small molecules, clusters and solids started near equilibrium, and no flexible molecule started far from a minimum; the relaxers are built for the first regime, and Sella and the optimizers built on it for the second.\par}

\clearpage
\phantomsection\label{si:cost}
\noindent Supplementary Table 7: \textbf{Compute and output tokens per run and configuration.} Unit-runs and output tokens as defined in Methods (Compute accounting). Transcripts are missing for some early runs, so the token column gives the number of runs it covers.

\needspace{8\baselineskip}\medskip\noindent \textbf{Kernels} (wall time), per run:

\smallskip
{\scriptsize
\begin{tabularx}{\linewidth}{>{\hsize=1.913\hsize\raggedright\arraybackslash}X>{\hsize=0.696\hsize\raggedright\arraybackslash}X>{\hsize=0.696\hsize\raggedright\arraybackslash}X>{\hsize=0.696\hsize\raggedright\arraybackslash}X}
\toprule
\textbf{\hspace{0pt}} & \textbf{\hspace{0pt}run 1} & \textbf{\hspace{0pt}run 2} & \textbf{\hspace{0pt}run 3} \\ \midrule
\hspace{0pt}unit-runs (six tasks) & \hspace{0pt}5,709 & \hspace{0pt}4,115 & \hspace{0pt}28,919 \\
\hspace{0pt}output tokens (coordinator and kernel agents) & \hspace{0pt}0.31 M & \hspace{0pt}0.40 M & \hspace{0pt}1.59 M \\
\hspace{0pt}speedup with the agents' C disabled (geometric mean over kernels) & \hspace{0pt}23.9$\times$ & \hspace{0pt}24.1$\times$ & \hspace{0pt}21.8$\times$ \\
\bottomrule
\end{tabularx}
\normalsize}
\smallskip

\needspace{8\baselineskip}\medskip\noindent \textbf{Optimizers} (force calls), per configuration:

\smallskip
{\scriptsize\setlength{\tabcolsep}{4pt}
\begin{tabularx}{\linewidth}{>{\hsize=2.066\hsize\raggedright\arraybackslash}X>{\hsize=0.393\hsize\raggedright\arraybackslash}X>{\hsize=1.082\hsize\raggedright\arraybackslash}X>{\hsize=0.787\hsize\raggedright\arraybackslash}X>{\hsize=0.689\hsize\raggedright\arraybackslash}X>{\hsize=0.984\hsize\raggedright\arraybackslash}X}
\toprule
\textbf{\hspace{0pt}configuration} & \textbf{\hspace{0pt}runs} & \textbf{\hspace{0pt}iterations per run} & \textbf{\hspace{0pt}unit-runs, total} & \textbf{\hspace{0pt}unit-runs per run} & \textbf{\hspace{0pt}output tokens per run (runs with a transcript)} \\ \midrule
\hspace{0pt}Opus 5 single-phase, library start & \hspace{0pt}18 & \hspace{0pt}500 & \hspace{0pt}425,094 & \hspace{0pt}23.6 k & \hspace{0pt}0.31 M (11 of 18) \\
\hspace{0pt}Opus 5 two-phase, library start & \hspace{0pt}18 & \hspace{0pt}500 (one run 561) & \hspace{0pt}413,627 & \hspace{0pt}23.0 k & \hspace{0pt}0.41 M (12 of 18) \\
\hspace{0pt}Opus 5 parameter floor & \hspace{0pt}8 & \hspace{0pt}$\leq$ 100 & \hspace{0pt}32,614 & \hspace{0pt}4.1 k & \hspace{0pt}0.10 M (2 of 8) \\
\hspace{0pt}Opus 5 single-phase, from scratch & \hspace{0pt}6 & \hspace{0pt}500 & \hspace{0pt}126,693 & \hspace{0pt}21.1 k & \hspace{0pt}0.43 M (4 of 6) \\
\hspace{0pt}Opus 5 two-phase, from scratch & \hspace{0pt}6 & \hspace{0pt}500 & \hspace{0pt}118,666 & \hspace{0pt}19.8 k & \hspace{0pt}0.46 M (6 of 6) \\
\hspace{0pt}Sonnet 5 single-phase, from scratch & \hspace{0pt}6 & \hspace{0pt}500 & \hspace{0pt}147,620 & \hspace{0pt}24.6 k & \hspace{0pt}2.54 M (6 of 6) \\
\hspace{0pt}Fable 5.1 single-phase, from scratch & \hspace{0pt}6 & \hspace{0pt}500 & \hspace{0pt}192,338 & \hspace{0pt}32.1 k & \hspace{0pt}0.55 M (6 of 6) \\
\hspace{0pt}Opus 5 informed & \hspace{0pt}6 & \hspace{0pt}100 & \hspace{0pt}39,113 & \hspace{0pt}6.5 k & \hspace{0pt}0.21 M (6 of 6) \\
\hspace{0pt}Fable 5.1 informed & \hspace{0pt}6 & \hspace{0pt}100 & \hspace{0pt}104,502 & \hspace{0pt}17.4 k & \hspace{0pt}0.35 M (6 of 6) \\
\hspace{0pt}\textbf{total} & \hspace{0pt}\textbf{80} & \hspace{0pt} & \hspace{0pt}\textbf{1,600,267} & \hspace{0pt} & \hspace{0pt} \\
\bottomrule
\end{tabularx}
\normalsize}
\smallskip

{\footnotesize\begin{itemize}\setlength{\itemsep}{1pt}
\item One Sonnet 5 relaxer run (run 3) used 187 worker subagents and 7.21 M output tokens, nearly half of the Sonnet 5 total.
\item Wall time is not compared: it depends on the algorithm being optimized (benchmarks get cheaper as it improves) and on operator interventions.
\end{itemize}}

\clearpage
\begin{figure}[h]
\centering
\includegraphics[width=\linewidth]{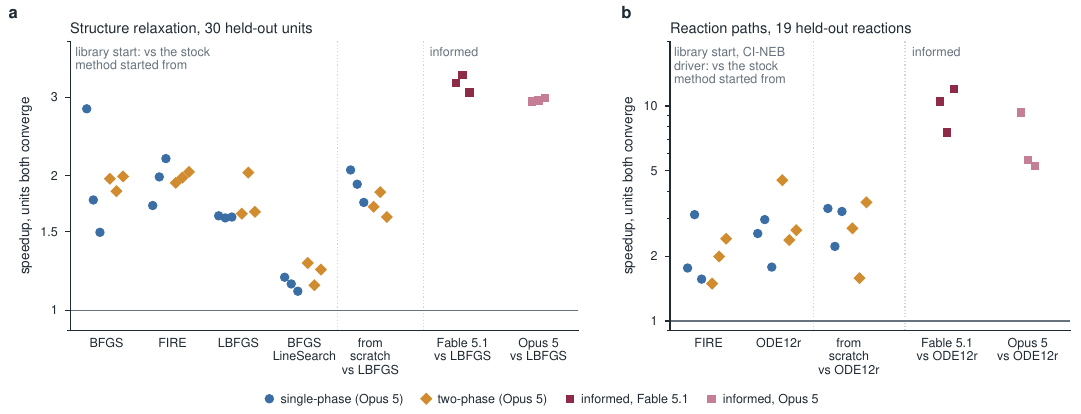}
\caption{\textbf{Held-out speedups without the penalty.} Each run's speedup over its reference (reference mean / run mean) on the held-out units that both converge, without penalty, for the runs of Fig. 3: \textbf{a}, structure relaxation, 30 units; \textbf{b}, reaction paths, 19 reactions. Log scale; dotted lines separate the starting points. References: library start, the stock method started from; from scratch and informed, L-BFGS (relaxation) and CI-NEB driven by ODE12r (paths). Where neither the run nor its reference leaves a unit unconverged, as for all but two relaxation runs, the value equals the speedup under the pre-specified metric. Grey line, speedup 1. Marker shapes as in Fig. 3.}
\label{fig:Snopen}
\end{figure}

\end{document}